\documentclass{aa}  

\usepackage{graphicx}
\usepackage{txfonts}
\usepackage{lipsum}
\usepackage{subcaption}         

\usepackage{lscape}             

\usepackage{placeins}

\usepackage{textgreek}
\usepackage{overpic}
\usepackage{xcolor}

\usepackage{hyperref}

\begin{document}

   \title{
   The pulsar population of Omega Centauri: New TRAPUM searches, timing, and cluster dynamics with MeerKAT and Parkes}

   \author{Miquel Colom i Bernadich\inst{1,2}
        \and Shi Dai\inst{3,4}
        \and Federico Abbate\inst{1}
        \and Matthew Kerr\inst{5}
        \and Matteo Bachetti\inst{1}
        \and Yash Bhargava\inst{1}
        \and Sarah Buchner \inst{6}
        \and Simon Johnston \inst{3}
        \and Marta Burgay \inst{1}
        \and Andrea Possenti\inst{1}
        \and Rouhin Nag\inst{1,7}
        \and Alessandro Ridolfi\inst{8}
        \and Amodio Carleo\inst{1}
        \and Alessandro Corongiu\inst{1}
        \and Paulo C. C. Freire\inst{2}
        \and Fernando Camilo\inst{6}
        \and Weiwei Chen\inst{2}
        \and Mario Cadelano\inst{9,10}
        \and Dhanraj Risbud \inst{2,8}
        \and Prajwal V. Padmanabh \inst{11,12}
        \and David J. Champion \inst{2}
        \and Michael Kramer \inst{2}
        \and Benjamin Stappers \inst{17}
        \and Maciej Serylak \inst{13}
        \and Vishnu Balakrishnan \inst{14}
        \and Matthew Bailes \inst{15,16}
        \and Arunima Dutta \inst{2}
        \and Laila Vleeschower\inst{17,18}
        \and Vivek Venkatraman Krishnan \inst{2}
        \and Yunpeng Men \inst{2}
        \and Ewan Barr \inst{2}
        }

   \institute{INAF -- Osservatorio Astronomico di Cagliari, via della Scienza 5, 09047 Selargius (CA), Italy
             \and
             Max-Planck-Institut für Radioastronomie, Auf dem Hügel 69, D-53121 Bonn, Germany\\
             \email{miquel.colomibernadich@inaf.it, mcbernadich@mpifr-bonn.mpg.de}
             \and
             Australia Telescope National Facility, CSIRO, Space and Astronomy, PO Box 76, Epping NSW 1710, Australia
             \and
             School of Science, Western Sydney University, Locked Bag 1797, Penrith, NSW 2751, Australia
             \and
             Space Science Division, Naval Research Laboratory, Washington, DC 20375–5352, USA
             \and
             South African Radio Astronomy Observatory, Liesbeek House, River Park, Cape Town 7705, South Africa
             \and
             University of Cagliari, Monserrato University Campus - SP Monserrato-Sestu Km 0,700 - 09042 Monserrato (CA), Italy
             \and Fakultät für Physik, Universität Bielefeld, Postfach 100131, D-33501 Bielefeld, Germany
             \and Department of Physics and Astronomy ‘Augusto Righi’, University of Bologna, via Gobetti 93/2, I-40129 Bologna, Italy
             \and
             INAF – Astrophysics and Space Science Observatory of Bologna, via Gobetti 93/3, I-40129 Bologna, Italy
             \and
             Max Planck Institute for Gravitational Physics (Albert Einstein Institute), D-30167 Hannover, Germany
             \and
             Leibniz Universit\"{a}t Hannover, D-30167 Hannover, Germany
             \and
             SKA Observatory, Jodrell Bank, Lower Withington, Macclesfield, Cheshire, SK11 9FT, UK
             \and
             Center for Astrophysics, Harvard \& Smithsonian, Cambridge, MA 02138-1516, USA
             \and
             Centre for Astrophysics and Supercomputing, Swinburne University of Technology, PO Box 218, Hawthorn, VIC 3122, Australia
             \and
             OzGrav: The ARC Centre of Excellence for Gravitational Wave Discovery, Hawthorn, VIC 3122, Australia
             \and
             Jodrell Bank Centre for Astrophysics, Dept. of Physics \& Astronomy, The University of Manchester, Manchester M13 9PL, UK 
             \and
             Center for Gravitation, Cosmology, \& Astrophysics, Dept. of Phys., Univ. of Wisconsin-Milwaukee, Milwaukee, WI 53201, USA
             }

   \date{Submitted March 03, 2026; accepted October 10, 2026}
 
  \abstract
  {Millisecond pulsars (MSPs) are probes of globular clusters (GCs), tracing their stellar evolution, cluster dynamics, and gravitational potential. We searched for and timed MSPs in the GC Omega Centauri, then we investigated their high-energy emission, the cluster potential, and MSP demographics. We performed Fourier-domain acceleration and jerk searches on MeerKAT observations, followed by pulsar timing using MeerKAT and Parkes Murriyang data spanning 2021--2025. We discovered the new isolated MSP PSR J1326$-$4728S (hereafter S), with a spin period of 4.538 ms and a dispersion measure of 96.24 pc\,cm$^{-3}$. We updated the timing solutions for pulsars A--E, obtained the first timing solutions for G, H, and K, and updated the orbital parameters for I, L, N, and Q. The latter three, I, N, and Q, have significantly longer orbital periods ($P_\mathrm{b}=1$--7\,days) than previously reported. Pulsars B, G, H, K, and L are black widows ($P_\mathrm{b}=2$--4\,hours); I is a wide-orbit ($P_\mathrm{b}=3.50$\,days) MSP with a light $\sim$0.05\,M$_\odot$ companion; N and Q have $\sim$0.25\,M$_\odot$ companions; and N shows a significant orbital eccentricity ($e = 0.093$). We folded Fermi-LAT and NICER photons to search for high-energy pulsations, but failed to detect them from individual MSPs. Significant spin period derivatives were measured for eight MSPs and interpreted as arising from the cluster gravitational potential. The inferred line-of-sight accelerations are consistent with a King-model gravitational potential, and allow us to set an upper central black hole mass limit of $M_\mathrm{BH}<8\times10^{4}$\,M$_\odot$ at 90\% confidence. The high fraction of isolated MSPs (58\%) and black widow systems (63\% of binaries), and possibly the eccentricity of N, are difficult to reconcile with MSP population predictions based solely on encounter rates. Instead, these properties require alternative formation channels or reflect the complex evolutionary history of Omega Centauri, with part of its MSP population having formed in denser environments than those observed today.}


   \titlerunning{The pulsar population of Omega Centauri with MeerKAT and Parkes}

   \maketitle
   \nolinenumbers

\section{Introduction}

\begin{table*}[h!]
\caption[]{\label{summary_table} The 19 known pulsars in \textomega~Cen and their parameters.
}
\centering
\begin{tabular}{lcccccccccc}
\hline
\hline \\[-1.5ex]
PSR & Discovery & $\theta$ & $P_\mathrm{s}$ & $\dot P_\mathrm{s}$ & DM & $P_\mathrm{b}$ & $x$ & $e$ & $M_\mathrm{c,min}$ & $M_\mathrm{c,med}$ \\
 &  & \arcmin & ms & $10^{-20}$ s\,s$^{-1}$ & pc\,cm$^{-3}$ & days & ls &  & M$_\odot$  & M$_\odot$ \\
\hline \\[-1.5ex]
A$^\ast$ &  \citet{dai2020omcen} &      1.93$^\mathrm{ a}$ & 4.109 &  $2.73$ & 100.33 & \multicolumn{4}{c}{isolated} \\
B$^\ast$ &  \citet{dai2020omcen} &     0.76$^\mathrm{ a}$ & 4.792 & $-5.43$ & 100.28 & 0.0896 & 0.0215 & $\lesssim10^{-4}$ & $0.013$ & $0.016$ \\
C$^\ast$ &  \citet{dai2020omcen} &     1.98$^\mathrm{ a}$ & 6.868 &  $1.01$ & 100.66 & \multicolumn{4}{c}{isolated} \\
D$^\ast$ &  \citet{dai2020omcen} &     2.50$^\mathrm{ a}$ & 4.579 & $-4.12$ &  96.55 & \multicolumn{4}{c}{isolated} \\
E$^\ast$ &  \citet{dai2020omcen} &     1.58$^\mathrm{ a}$ & 4.208 &  $1.63$ &  94.34 & \multicolumn{4}{c}{isolated} \\
F & \citet{chen2023omcen} & $\sim$1.0$^\mathrm{ c}$ & 2.273 &     ... &   98.29 & \multicolumn{4}{c}{isolated} \\
G$^\ast$ & \citet{chen2023omcen} &     1.96$^\mathrm{ a}$ & 3.304 &  $2.77$ &  99.75 & 0.1086 & 0.0322 & $\lesssim10^{-4}$ & $0.018$ & $0.021$ \\
H$^\ast$ & \citet{chen2023omcen} &     0.56$^\mathrm{ a}$ & 2.520 &  $3.99$ &  98.17 & 0.1357 & 0.0219 & $\lesssim10^{-4}$ & $0.010$ & $0.012$\\
I$^\dagger$ & \citet{chen2023omcen} & 3.53$^\mathrm{ b}$ & 18.95 &     ... & 102.56 & 3.5083 & 0.7852 & $\lesssim10^{-3}$ & $0.043$ & $0.050$ \\
J & \citet{chen2023omcen} & 1.80$^\mathrm{ b}$ & 3.686 &     ... &  97.28 & \multicolumn{4}{c}{isolated} \\
K$^\ast$ & \citet{chen2023omcen} &     1.89$^\mathrm{ a}$ & 4.716 & $-0.91$ &  94.78 & 0.0939 & 0.0680 & $\lesssim10^{-4}$ & $0.042$ & $0.049$ \\
L$^\dagger$  & \citet{chen2023omcen} & 3.32$^\mathrm{ b}$ & 3.537 &     ... & 101.48 & 0.1589 & 0.0618 & $\lesssim10^{-4}$ & $0.027$ & $0.031$ \\
M & \citet{chen2023omcen} & $\sim$2.4$^\mathrm{ c}$ & 4.604 &     ... & 101.47 & \multicolumn{4}{c}{isolated} \\
N$^\dagger$ & \citet{chen2023omcen} & 2.66$^\mathrm{ b}$ & 6.884 &     ... &   102.1 & 6.3566 & 5.7231 &           0.0928 & $0.232$ & $0.272$ \\
O & \citet{chen2023omcen} & $\sim$1.5$^\mathrm{ c}$ & 6.160 &     ... &  94.305 & \multicolumn{4}{c}{isolated} \\
P & \citet{chen2023omcen} & $\sim$1.0$^\mathrm{ c}$ & 2.795 &     ... & 102.17 & \multicolumn{4}{c}{isolated} \\
Q$^\dagger$ & \citet{chen2023omcen} & $\sim$2.3$^\mathrm{ c}$ & 4.130 &     ... &  95.955 & 1.5400 & 1.9938 & $\lesssim10^{-4}$ & $0.205$ & $0.241$ \\
R & \citet{chen2023omcen} & $\sim$3.9$^\mathrm{ c}$ & 10.29 &     ... &   102.2 & \multicolumn{4}{c}{isolated} \\
S &             this work & 2.32$^\mathrm{ d}$ & 4.538 &     ... &  96.24 & \multicolumn{4}{c}{isolated} \\
\hline
\hline
\end{tabular}

\tablefoot{Listed parameters: angular separation from the cluster centre, $\theta$; spin period, $P_\mathrm{ms}$; spin period derivative, $\dot P_\mathrm{s}$; dispersion measure, DM; orbital period, $P_\mathrm{b}$; orbital eccentricity, $e$; and minimum companion mass, $M_\mathrm{c}$. Here, $\theta$ assumes $\textrm{RAJ},\textrm{DECJ}=13^\textrm{h}26^\textrm{m}47.24^\textrm{s},-47$:28:46.5 for the cluster \citep{harris2010catalog}. Minimum and median companion masses ($M_\textrm{c,min}$, $M_\textrm{c,med}$) were derived from the mass function (pulsar mass $M_\mathrm{p}=1.35$\,$M_\odot$, inclination angle $i=90,\,60^\circ$).\\
$^\ast$Parameters for this pulsar measured in phase-connected timing models derived in this work. $^\dagger$Parameters for this pulsar measured in orbital models derived in this work.\\
$^\mathrm{a}$Timing position updated/measured in this work. $^\mathrm{b}$\texttt{SeeKAT}-localised position from multiple MeerKAT/APSUSE beam detections, reported in \cite{chen2023omcen}. $^\mathrm{c}$Position from single MeerKAT/APSUSE beam detection, reported in \cite{chen2023omcen}. $^\mathrm{d}$\texttt{SeeKAT}-localised position from multiple MeerKAT/APSUSE beam detections, derived in this work.}

\end{table*}

Omega Centauri (\textomega~Cen, also NGC~5139) is a unique globular cluster (GC). With a mass of $3.55\times10^{6}$\,M$_\odot$ and a core radius of 4.30\,pc \citep{baumgardt2018catalogueGCs}, it is the largest and most massive GC in the Milky Way. It follows a retrograde orbit \citep{dinescu1999velocities} and exhibits exceptional stellar complexity, with spatially mixed metallicity populations \citep{nitschai2024mixing}, 15 distinct main-sequence populations \citep{bellini2017populations}, and a stellar age spread of 10--13 Gyr \citep{vilanova2014metallicityage,tailo2016mosaic,clontz2024agespread}. These properties, among others, suggest that \textomega~Cen is the stripped core of a dwarf galaxy accreted by the Milky Way \citep{lee1999merger,hilker2000dwarf,bekki2003formation,pagnini2025merger}, linked to the Sequoia \citep{muon2019squoia} or Gaia-Enceladus mergers \citep{massari2019origin,pfeffer2021accreted,limberg2022reconstructiong}, and Galactic halo tidal stellar streams \citep{majweski2012stream,ibata2019stream}.

A related debate concerns the possible presence of a central intermediate-mass black hole (IMBH) in \textomega~Cen. Some studies support its existence, suggesting a mass in the range of $10^{3}$–$10^{5}$\,M$_\odot$ from stellar kinematics \citep{noyola2008imbh,noyola2010imbh,jalali2012dynamical,baumgardt2017imbhs,haberle2024inmbh}. In contrast, other studies place upper limits of $\lesssim$\;10$^{4}$\,M$_\odot$ \citep{vanderMarel2010limits,banares2025omcenconstraints}, or instead favour an extended central cluster \citep{zocchi2017anisotropy,zocchi2019kinematics,evans2022components,banares2025omcenconstraints}.

Pulsars are powerful probes of GC environments due to their timing stability and sensitivity to cluster properties. The observed spin period derivative ($\dot P_\mathrm{s}$) includes both intrinsic spin-down and kinematic contributions, the latter arising from line-of-sight (LOS) acceleration in the cluster’s gravitational potential. Therefore, their measurements constrain the cluster acceleration field and thus its mass distribution \citep[e.g.][]{prager2017timingter5,freire2017tuc47}. In addition, MSP populations also trace stellar dynamics, with total and isolated populations possibly correlating with the total stellar encounter rate \citep{verbunt1987origin,hui2010ecounters,bahramian2013encounters} and the encounter rate per binary \citep{verbunt2014disruption,raiso1995eccentricities}. Finally, MSP populations are linked to the global $\gamma$-ray emission of GCs \citep{deMenezes2019gammaray}, with $\gamma$-ray pulsations being detectable after folding with radio timing solutions \citep[e.g.][]{freire2011detection}.

In this study, we investigated the pulsar population of \textomega~Cen with MeerKAT\footnote{\url{https://www.sarao.ac.za/science/meerkat/}} and Parkes\footnote{\url{https://www.parkes.atnf.csiro.au/}}. Our work is a continuation of \cite{dai2020omcen,dai2023omcen} and \cite{chen2023omcen}, which reported previous pulsar discoveries and timing efforts in this cluster with the same instruments. This study is part of the TRAnsients and PUlsars with MeerKAT (TRAPUM\footnote{\url{https://www.trapum.org/}} \citealt{stappers2016trapum}) globular cluster survey, the MeerTIME\footnote{\url{http://www.meertime.org/}} \citep{bailes2020facility} project, and the P1041 Parkes observing program.

Several questions remain unresolved regarding \textomega-Cen's pulsar population. Prior to this work, 18 millisecond pulsars (MSPs) had been reported in \textomega~Cen (see Table~\ref{summary_table}). These include eight binaries and ten isolated pulsars. In  \cite{dai2020omcen}, the discoveries of PSR J1326$-$4728A to E (hereafter A--E) with Parkes were reported, and in \cite{dai2023omcen} timing solutions were derived. Furthermore, in \cite{chen2023omcen}, the discoveries of F--R with MeerKAT were reported.

\cite{dai2023omcen} used the $\dot P_\mathrm{s}$ measurements of A--E to place constraints on the cluster acceleration field, showing broad consistency with the GC density profile and potential presented by \citet{king1962profile}. Possible exceptions were B and D, whose inferred LOS accelerations marginally exceeded the expected limits. We aimed to derive new timing solutions to further constrain the mass distribution of \textomega~Cen and its putative central IMBH. 

Furthermore, \textomega~Cen hosts a high isolated pulsar fraction ($\sim$55\%), posing a challenge to our understanding of pulsar population scenarios. Isolated MSPs in GCs are thought to arise from binary disruptions during stellar encounters \citep{verbunt2014disruption}. However, as discussed in \citet{chen2023omcen}, this interpretation is problematic for \textomega~Cen given its low encounter rate per binary. For instance, \textomega~Cen has an encounter rate per binary more than ten times lower than that of Terzan 5 and 47 Tucanae, which instead host isolated fractions of only $\sim$40\% \citep{padmanabh2024ter5,freire2017tuc47}. This could be directly correlated with an excess of X-ray sources observed in \textomega~Cen and other clusters with low encounter rates, suggesting that compact binaries might form and evolve via alternative mechanisms in low-density environments \citep{deMenezes2023dynamics}. Therefore, our aim was to perform pulsar searches using new TRAPUM observations to refine the population statistics.

Finally, \textomega~Cen coincides with a point-like $\gamma$-ray source of $L_\textrm{\textgamma}=(3.6\pm0.3)\times10^{34}$\,erg\,s$^{-1}$ \citep{abdo2010fermi}, whose origin remains uncertain. \cite{deMenezes2019gammaray} estimates that the $\gamma$-ray luminosity is consistent with a population of 10 to 30 MSPs. Alternatively, it has been postulated that the $\gamma$-ray emission originates from dark matter annihilation \citep{brown2019annihilation,reynoso-cordova2021darkmatter}. \citet{dai2023omcen} searched for $\gamma$-ray pulsations using the timing solutions of A--E, but without success. The non-detection was interpreted as the emission being distributed over the collective population. However, new discoveries provide a renewed opportunity for $\gamma$-ray pulsation searches. 

In this paper, we address these topics through new radio pulsar searches and timing analyses with Parkes and MeerKAT observations (details in Section~\ref{observations_section}). In particular, we performed new pulsar searches in previously unsearched beams of earlier TRAPUM \textomega~Cen observations with the MeerKAT L-band receiver (1.284\,GHz). These beams target radio sources identified with Australia Telescope Compact Array \citep[ATCA, see][]{dai2020omcen}. The nature of these sources is currently unknown and they might potentially be associated with pulsar emission. For this reason, they were included in our search target list. We also searched a new TRAPUM observation with the MeerKAT UHF receiver (0.816\,GHz) to identify pulsars with potentially steep spectral indices that may have been missed in the L-band searches reported by \citet{chen2023omcen}. These searches, along with the newly discovered pulsar S, are reported in Section~\ref{searches_section}.

Using both MeerKAT and Parkes observations, we updated the timing solutions for A--E, provided first-time timing solutions for G, H, and K, and updated the orbital parameters of I, L, N, and Q. These results are reported in Section~\ref{timing_section}. We also performed renewed searches for $\gamma$-ray and X-ray pulsations using Fermi-LAT\footnote{\url{https://science.nasa.gov/mission/fermi/}} data \citep[Large Area Telescope;][]{Atwood09}, NICER\footnote{\url{https://science.nasa.gov/mission/nicer/}} observations \citep{gendreau2012nicer}, which we report in Section~\ref{high_section}. In addition, we also explored possible identifications in XMM-Newton\footnote{\url{https://www.cosmos.esa.int/web/xmm-newton}} data. In Section~\ref{dynamics_section}, we present updated constraints on the cluster’s mass distribution using $\dot{P}_\mathrm{s}$ measurements obtained from our timing solutions. Based on these measurements, we placed an upper limit on the mass of the central IMBH of $M_\mathrm{BH}<8 \times 10^4$\,$M_\odot$ at 90\% confidence based solely on timing measurements. Finally, in Section~\ref{population_section} we discuss the potential origin of the MSP population in the context of the cluster’s encounter rates.

\section{Observations included in this study}\label{observations_section}

\subsection{Parkes observations}

Parkes observations performed with the Ultra Wideband Low-frequency receiver of the Murriyang telescope \citep[UWL, 704–4032 MHz;][]{hobbs2020uwl} were used exclusively for the timing analysis. They were obtained as part of the long-term P1041 observing project, a continuation of \cite{dai2020omcen,dai2023omcen} dedicated to the follow-up of pulsars in \textomega~Cen. In this study, we included all observations carried out between 1 March 2020 and 20 May 2025, performed at approximately monthly cadence with a break between February 2024 and April 2025.

Observations were recorded in search-mode with the MEDUSA backend \citep{hobbs2020uwl} with full-Stokes resolution, 3328 frequency channels, coherent de-dispersion at the average value of \textomega~Cen pulsars ($\mathrm{DM}=99$\,pc\,cm$^{-3}$), and with 64\,\textmu s sampling. Individual observing scans lasted between three and five hours, accumulating a total of 160 hours of observing time. In addition, before the main observations, Murriyang also recorded 3-min long observations of the noise diode for polarisation calibration. With Murriyang targeting the centre of the cluster, the antenna beam ($\sim$30$^\prime$) provided good coverage of all pulsars simultaneously.

\subsection{MeerKAT observations}\label{meerkat_observations}

MeerKAT data used in this work were collected on eight observing sessions within 2021--2025 as part of the TRAPUM GC and MeerTIME projects (see Table~\ref{observations_table} for a listing). These data were used for both searching and timing. The observations were beamformed by the Filterbanking Beamformer User Supplied Equipment (FBFUSE), and were subsequently stored as search-mode filterbanks in the Accelerated Pulsar Search User Supplied Equipment \citep[APSUSE, see][]{ebarr2018sband,padmanabh2023mmgps}. APSUSE data offers only total intensity, with no polarimetry information nor coherent de-dispersion.

The main objective of TRAPUM observations is new pulsar searches. Its observations implement a full coherent beam tiling derived with MOSAIC\footnote{\url{https://github.com/wchenastro/mosaic}} \citep{chen2021mosaic}, with additional coherent beams synthesised on optimal pulsar positions as reported by \citet{dai2023omcen} and \citet{chen2023omcen}, and ATCA source positions as reported in \citet{dai2020omcen} (see Table~\ref{observations_table} in Appendix~\ref{observations_appendix} for a more detailed listing). On the other hand, MeerTIME observations were of follow-up nature, and data stored in APSUSE includes only beams synthesised at known pulsars and ATCA source positions.

All observations listed in Table~\ref{observations_table} were used in the timing efforts by folding the coherent beams synthesised on the pulsars, or the nearest beams in their absence. In addition, the beams centred on ATCA sources from the 21 March 2021 L-band observation, and the MOSAIC tiling beams from the 11 February 2025 observation were also searched for new pulsars.

In addition to the APSUSE data described above, further MeerKAT data were obtained as part of the MeerTIME programme using the Pulsar Timing User Supplied Equipment (PTUSE) backend \citep{bailes2020facility}. Many of these data were obtained simultaneously with the APSUSE data. In this work, we use only the 14 February 2024 L-band PTUSE beam, which targeted pulsar L exclusively.

\section{New searches on TRAPUM data}\label{searches_section}

\subsection{Searched beams}

We used the APSUSE data to search for new pulsars. The objectives were (i) to perform a deep search of the 12 coherent beams assigned to ATCA-imaged sources reported by \citet{dai2020omcen}, and (ii) to find new pulsars in a new 3-hour UHF observation.

The search of the ATCA beams was performed on the 21 March 2021 L-band observation. An exhaustive search of the MOSAIC coherent beam tiling of this dataset, with a sensitivity of 10\,\textmu Jy, was previously carried out by \citet{chen2023omcen}, resulting in the discovery of 13 new pulsars. However, that search did not include the beams centred on the ATCA sources, which would provide the strongest response to any pulsar signal associated with them. We complemented the work of \citet{chen2023omcen} by specifically searching the 12 remaining coherent beams assigned to the ATCA sources, whose locations are highlighted in orange in Fig.~\ref{beams_map}.

The search on new UHF data was performed on an 11 February 2025 observation, following a similar strategy as the search on the 21 March 2021 search presented by \citet{chen2023omcen}. We searched the beams of the MOSAIC tiling, which consisted of 248 coherent beams centred on the cluster and shown in grey in Fig.~\ref{beams_map}. At the time of submission, we had searched 66 beams in the central region of the tiling, highlighted in blue in Fig.~\ref{beams_map}.

Assuming a signal-to-noise (S/N) ratio detection threshold of 8 and a pulse duty cycle of 10\%, we estimate a sensitivity of 20\,\textmu Jy using the radiometer equation \citep{dewey1985search} and the MeerKAT specifications\footnote{\url{https://skaafrica.atlassian.net/wiki/spaces/ESDKB/pages/277315585/MeerKAT+specifications}} (details are provided in Appendix~\ref{sky_appendix}). The reduced sensitivity compared to the \citet{chen2023omcen} search arises from the shorter integration time, the smaller bandwidth of the UHF receiver, and the slightly higher system temperature at UHF relative to the L band. However, this is partially compensated by the increased pulsar flux density at lower frequencies: for pulsars with steep spectral indices ($<-2.0$), the flux density at the lower end of the UHF band (544\,MHz) can be up to three times higher than at L-band frequencies (816\,MHz).

As listed in Table~\ref{observations_table}, the March 2021 beams were recorded with a time sampling of 153.121\,\textmu s and 2048 frequency channels, which were subsequently sub-banded to 128 channels at the cluster DM. In contrast, the February 2025 beams were recorded with a time sampling of 120.471\,\textmu s and 4096 frequency channels, and later sub-banded to 256 channels.

\subsection{Implementation}

We implemented Fourier-domain jerk \citep{andersen2018jerk} and GPU-accelerated acceleration searches \citep{ransom2002fourier} on individual beams with \texttt{PRESTO}\footnote{\url{https://github.com/jintaoluo/presto2_on_gpu}}. The searches were run with \texttt{PULSAR\_MINER}\footnote{\url{https://github.com/alex88ridolfi/PULSAR_MINER}}, a \texttt{PRESTO}-based searching pipeline \citep{ridolfi2021miner}. Radio-frequency interference (RFI) was excised using \texttt{PRESTO}'s \texttt{rfifind}, masking 10 to 15\% of channels.

We adopted a DM search range of 90--110\,pc\,cm$^{-3}$, based on the DM of previously known pulsars. For the acceleration searches, we assumed a maximum Fourier-frequency derivative of $z_\mathrm{max}=200$, and allowed the summing of up to eight harmonics. For the jerk searches, we limited the Fourier-frequency derivatives to $z_\mathrm{max}=100$ and $w_\mathrm{max}=300$, respectively, and restricted the harmonic summing to four (see \citealt{andersen2018jerk} for the definition of these parameters). We also imposed a minimum FFT significance threshold of 3 for folding. Pulsar candidate plots folded with \texttt{prepfold} were inspected by eye.

Finally, to account for pulsars with extreme accelerations or jerks, we implemented segmented searches in addition to searches on the entire observations. The 4-hour files from the ATCA beams were segmented into chunks of 10, 30, 60, and 120 minutes, while the 3-hour files from the UHF observation were divided into segments of 10, 20, 45, and 90 minutes. Assuming that a pulsar signal can be recovered when the observation spans less than $\sim$10\% of the orbital period, this setup is theoretically sensitive to orbital periods as short as one hour.

The beams included in these searches are shown in the sky map presented in Fig.~\ref{beams_map} in Appendix~\ref{sky_appendix}. The search on ATCA beams is complete and yielded no new pulsars with flux densities above 10\,\textmu Jy at the L band. A new pulsar was discovered in the UHF MOSAIC tiling, on which we expand upon in the next section.

\subsection{UHF discovery: PSR J1326$-$4728S}\label{newPSR}

\begin{figure}[t!]
\centering
 \includegraphics[bb=275 80 685 430, width=\hsize, clip]{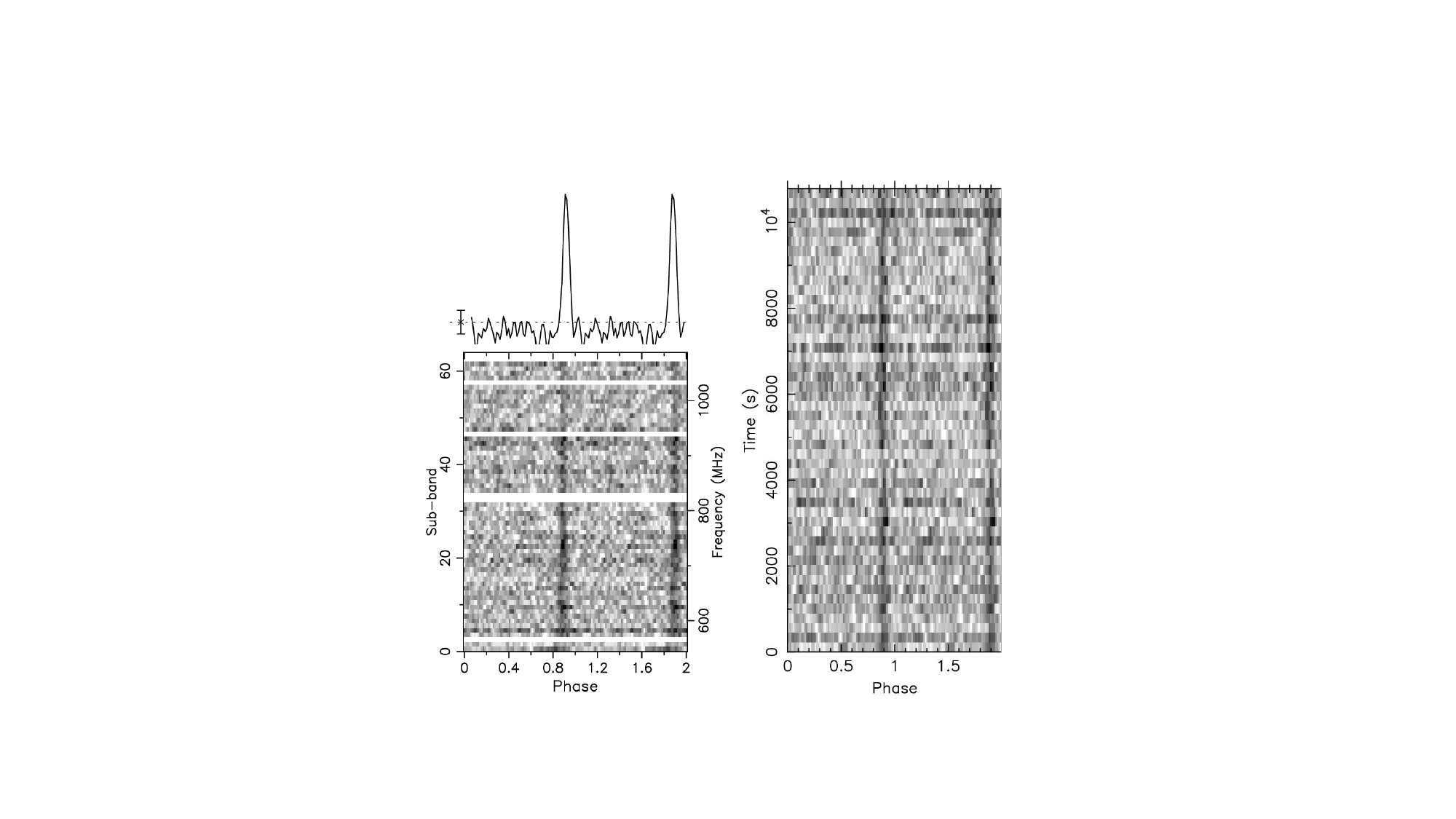}
 \caption{Discovery \texttt{prepfold} plots of pulsar S from the 2025 February 11 MeerKAT UHF observation (full three-hour integration). \textbf{Top-left:} Integrated pulse profile. \textbf{Bottom-left:} Frequency–phase greyscale. \textbf{Right:} Time–phase greyscale.}
 \label{S-discovery}
\end{figure}

Pulsar S was discovered in beam \texttt{cfbf00035} of the 11 February 2025 UHF observation. It was detected in the 45-min, 90-min and full integration searches, achieving a \texttt{prepfold} folded detection significance of 17.8\textsigma~in the full integration. It exhibited a barycentric spin period of $P_\mathrm{s}=4.537782062(58)$\,ms at a dispersion measure of $\mathrm{DM}=96.240$\,pc\,cm$^{-3}$, with no evidence for significant acceleration or jerk. Its integrated pulse profile, together with the frequency– and time–phase diagrams, is shown in Fig.~\ref{S-discovery}. Interpolating across bins, the half-height width of the pulse at UHF ($\sim$0.8\,GHz) is $310\pm70$\,\textmu s or $\sim$7\%, with the uncertainty arising from the coarse binning in spin phase.

\begin{figure}[t!]
\centering
 \begin{overpic}[bb=265 133 675 390, width=\hsize, clip]{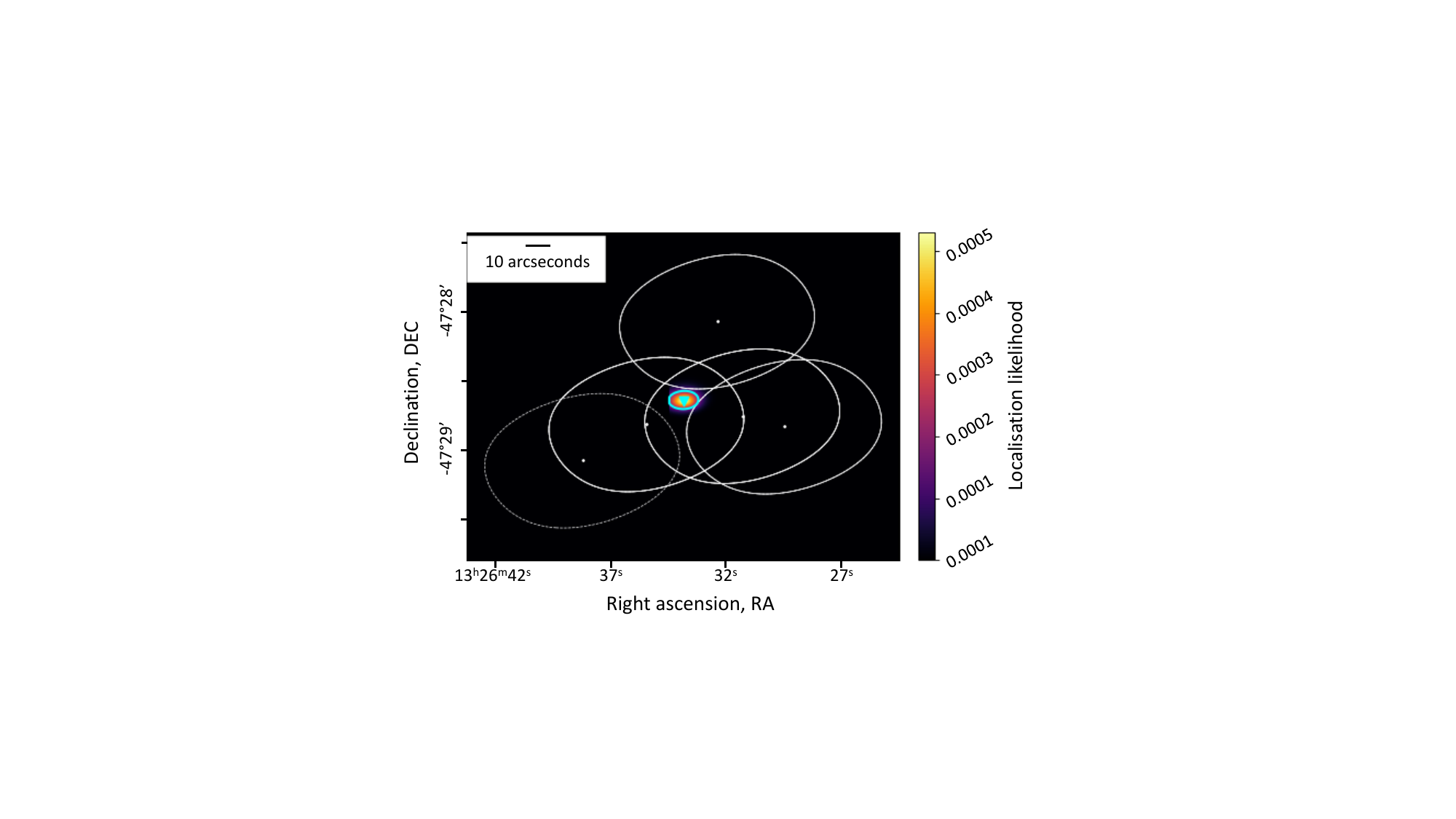}
 \put(29.7,23.5){\color{white}\small\texttt{cfbf00024}}
 \put(29.3,25.2){\color{white}\circle*{1}}
 \put(62.1,29){\color{white}\small\texttt{cfbf00037}}
 \put(61.7,30.7){\color{white}\circle*{1}}
 \put(55.5,32.5){\color{white}\small\texttt{cfbf00003}}
 \put(54.9,32.3){\color{white}\circle*{1}}
 \put(51.4,48){\color{white}\small\texttt{cfbf00051}}
 \put(50.9,47.6){\color{white}\circle*{1}}
 \put(40,29){\color{white}\small\bfseries\texttt{cfbf00035}}
 \put(39.5,31){\color{white}\circle*{1}}
 \end{overpic}
 \caption{\texttt{SeeKAT} localisation map of pulsar S, obtained from detections in multiple neighbouring MeerKAT/APSUSE beams. White ellipses indicate the 50\% effective gain boundaries of individual beams. Included beams and detection significances: \texttt{cfbf00003} (Pulsar D, 16.7\textsigma), \texttt{cfbf00024} (ATCA 7, 9.5\textsigma), \texttt{cfbf00037} (MOSAIC, 12.0\textsigma), \texttt{cfbf00035} (MOSAIC \& discovery, 17.8\textsigma), and \texttt{cfbf00051} (MOSAIC, 11.9\textsigma).}
 \label{S-localisation}
\end{figure}

Significant re-detections were achieved in the same UHF observation through direct re-folding of two neighbouring MOSAIC tiling beams, as well as the beams dedicated to pulsar D and ATCA source~7 from \citet{dai2020omcen} due to their proximity, resulting in a total of five significant detections (see Fig.~\ref{S-localisation}). We note that the detection at that UHF ATCA beam compared the non-detection in the L-band ATCA beam is due to the larger size of coherent beams at lower frequencies. We used \texttt{SeeKAT}\footnote{\url{https://github.com/BezuidenhoutMC/SeeKAT}} to perform a $\chi^2$ localisation of S based on the beam shapes, beam positions, and detection S/N \citep{bezuidenhout2023seekat}. The localisation, shown in Fig.~\ref{S-localisation}, yielded a refined position of 
$\textrm{RAJ},\textrm{DECJ}=13^{\mathrm{h}}26^{\mathrm{m}}33.76(1)^{\mathrm{s}}$, $-$47:28:38.5$\pm$1.5.

With the localised position, we folded the closest beams from the L-band observations obtained on 21 and 27 March 2021. No single beam was fully optimal, as the source lays close to the edges of all neighbouring beams. The resulting re-folded detections yielded a \texttt{prepfold} significance of $\sim$6\textsigma~over the full integration time, explaining why this pulsar was missed in \citet{chen2023omcen}. No change in the spin period was observed, indicating that pulsar S is likely not in binary.

Using the radiometer equation and assuming a pulse duty cycle of $\sim$7\%, we estimate a flux density of 50–60\,\textmu Jy at the UHF band. This estimate is uncertain, as it depends on the true position of the source relative to the beam centre. No polarimetric information for this pulsar is available due to the absence of PTUSE beams near its location in previous observations. The pulsar is not detected in the available Murriyang UWL observations upon folding, likely due to its faint nature.

\section{Timing analysis}\label{timing_section}

\subsection{Phase-connected timing solutions: Data analysis}

\begin{figure*}[h!]
\centering
 \includegraphics[trim=0 0 0 7, clip, width=0.99\hsize]{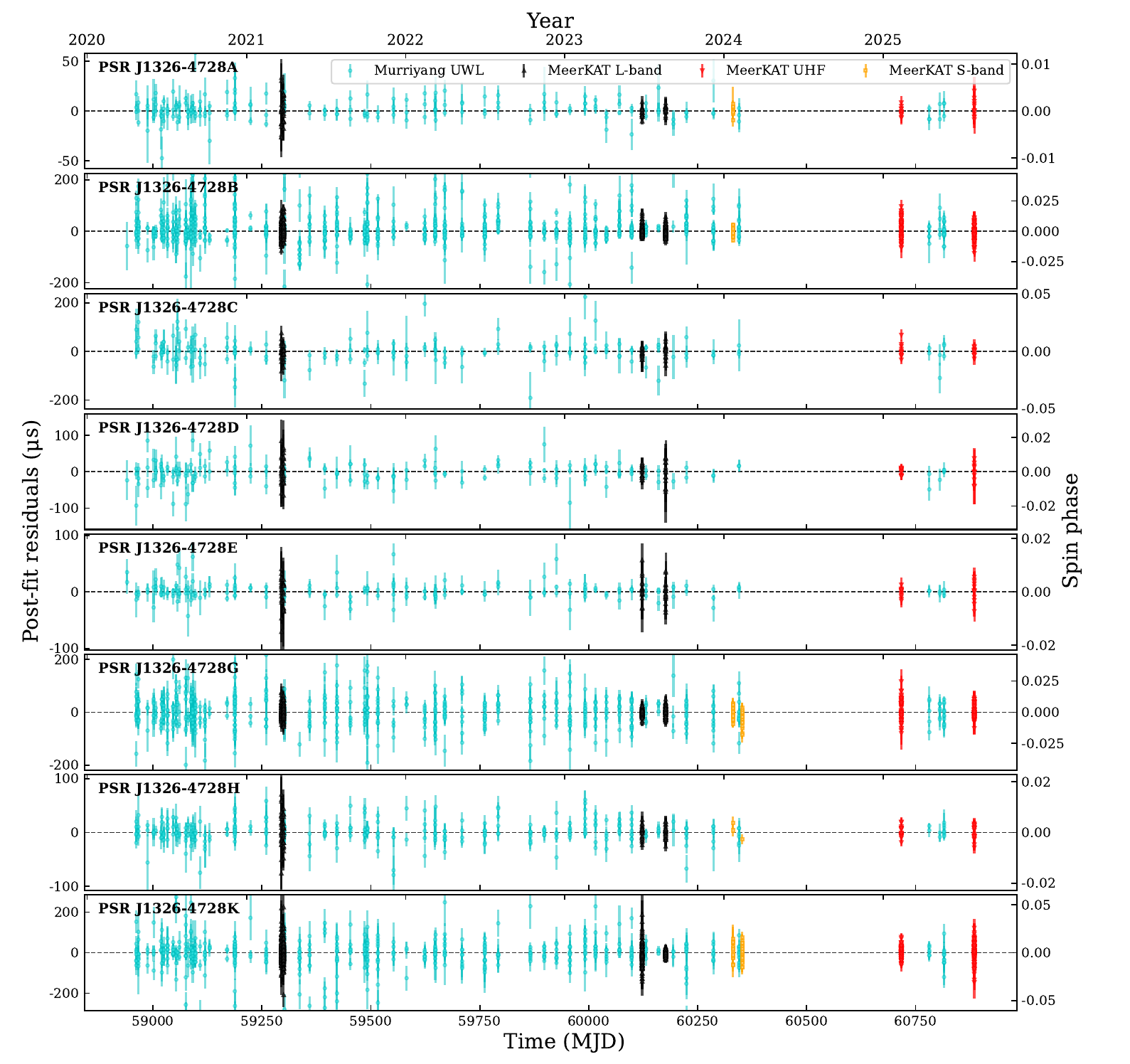}
 \caption{Combined MeerKAT and Murriyang timing residuals for A, B, C, D, E, G, H, and K as a function of time. A timing jump and DM offset are applied at the 2021 L-band, 2023 L-band, S-band, and UHF MeerKAT ToAs.}
 \label{timing_residuals}
\end{figure*}

We updated the phase-coherent timing solutions for A, B, C, D, and E, and derived timing solutions for G, H, and K for the first time. The solutions were obtained by fitting times of arrivals (ToAs) derived from MeerKAT and Murriyang observations. 

We folded the optimal MeerKAT filterbanks and the Murriyang UWL search-mode data using \texttt{dspsr}\footnote{\url{https://github.com/demorest/dspsr}} \citep{vanStraten2011}, and excised them of RFI with \texttt{clfd} \citep[an outlier-detection algorithm, ][]{morello2019}. We processed the folded pulsar archives with \texttt{PSRCHIVE} \footnote{\url{https://psrchive.sourceforge.net/}} \citep{hotan2004psrchive}, including the polarisation calibration of Murriyang UWL observations with \texttt{pac}. We performed our timing model fits with \texttt{TEMPO2}\footnote{\url{https://www.pulsarastronomy.net/pulsar/software/tempo2}} \citep{hobbs2006tempo2}.

To break the degeneracy between DM and spin evolution, we implemented a frequency-resolved timing strategy. MeerKAT L-band and UHF observations were reduced into two frequency sub-bands of an equal width, while Murriyang UWL observations were divided into three sub-bands: low (723--1093 MHz), mid (1094--1825 MHz), and high (1826--2656 MHz). Due to faintness at high frequencies, MeerKAT S-band ToAs were integrated over the full band.

Timing templates were derived using \texttt{PSRCHIVE}'s \texttt{paas}. For the MeerKAT L-band and UHF observations, we derived a template for each sub-band, producing frequency-resolved templates. For MeerKAT UHF, we used the time-integrated profiles from the 11 February 2025 observation to derive the template. For the L band, we treated the 2021 and 2023 observations as if they had been obtained with different backends, owing to differences in sub-banding and time resolution (see Table~\ref{observations_table}). These differences, combined with the lack of coherent de-dispersion, resulted in artificial phase and DM offsets when the observations were timed using the same template. We therefore derived one template from the time-integrated profiles of the two 2021 observations and another one from the time-integrated profiles of the 2023 observations. For the S band, we derived a single template from the frequency- and time-integrated profile of the February 2024 observations.

Murriyang UWL templates were generated for the individual low-, mid-, and high-frequency sub-bands stated above by integrating in time across all available observations to achieve a high S/N profile. These templates were then used as the frequency-resolved templates for the three UWL sub-bands. All of these processing steps were applied consistently to the eight pulsars timed in this section.

We derived the ToAs with the templates on the observations with \texttt{PSRCHIVE}'s \texttt{pat}. Tables~\ref{timing_isolated_appendix} and \ref{timing_binary_appendix} list the ToA integration times for each pulsar and backend. Typically, Murriyang UWL integration times are longer than in MeerKAT, and in the case of isolated pulsars we derived one Murriyang UWL ToA per observation and sub-band. MeerKAT S-band ToAs also required longer integration times than the L-band and UHF ToAs in some cases. We could not measure significant MeerKAT S-band ToAs for C, D, and E due to intrinsic faintness at radio frequencies larger than 2 GHz.

In the timing models, we accounted for the spin frequency and its first time derivative, $\nu$ and $\dot\nu$; the DM and its first time derivative, $\mathrm{DM}_0$ and $\mathrm{DM}_1$; along with the position and proper motion parameters RA, DEC, $\mu_\mathrm{RA}$, and $\mu_\mathrm{DEC}$. We included a fixed parallax parameter of $\mathrm{PX}=0.191$\,mas, consistent with the Gaia EDR3 measurement for \textomega~Cen \citep{soltis2021parallax}. Where applicable, we also included the second spin frequency derivative, $\ddot\nu$. For each MeerKAT observation group (L band 2021, L band 2023, S band, and UHF) we implemented an independent phase offset (timing jump) and a DM offset, using the Murriyang ToAs as reference.

For the binaries, we used the ELL1 timing model \citep{lange2001precision}, which implements the first and second Laplace-Lagrange $\eta=e\sin{\omega}$ and $\kappa=e\cos{\omega}$ (where $e$ and $\omega$ are the Keplerian eccentricity and angle of periastron parameters), as well as the time of the ascending node, $T_\mathrm{A}$, the projected semi-major axis, $x$, and the orbital period $P_\mathrm{b}$. We also included the orbital period derivative, $\dot P_\mathrm{b}$. As the starting point for the fits, we used the timing solutions provided in \citet{dai2023omcen} for A--E, along with the preliminary phase-connected solutions derived using our custom orbital fitting software\footnote{\url{https://github.com/mcbernadich/pulsar_orbit_solver}} and \texttt{dracula}\footnote{\url{https://github.com/pfreire163/Dracula}} for G, H, and K (Appendix~\ref{timing_appendix}).

\subsection{Phase-connected timing solutions: Results}\label{timing_results_section}

\begin{figure*}[h!]
\centering
 \includegraphics[trim=0 0 0 15, clip, width=0.99\hsize]{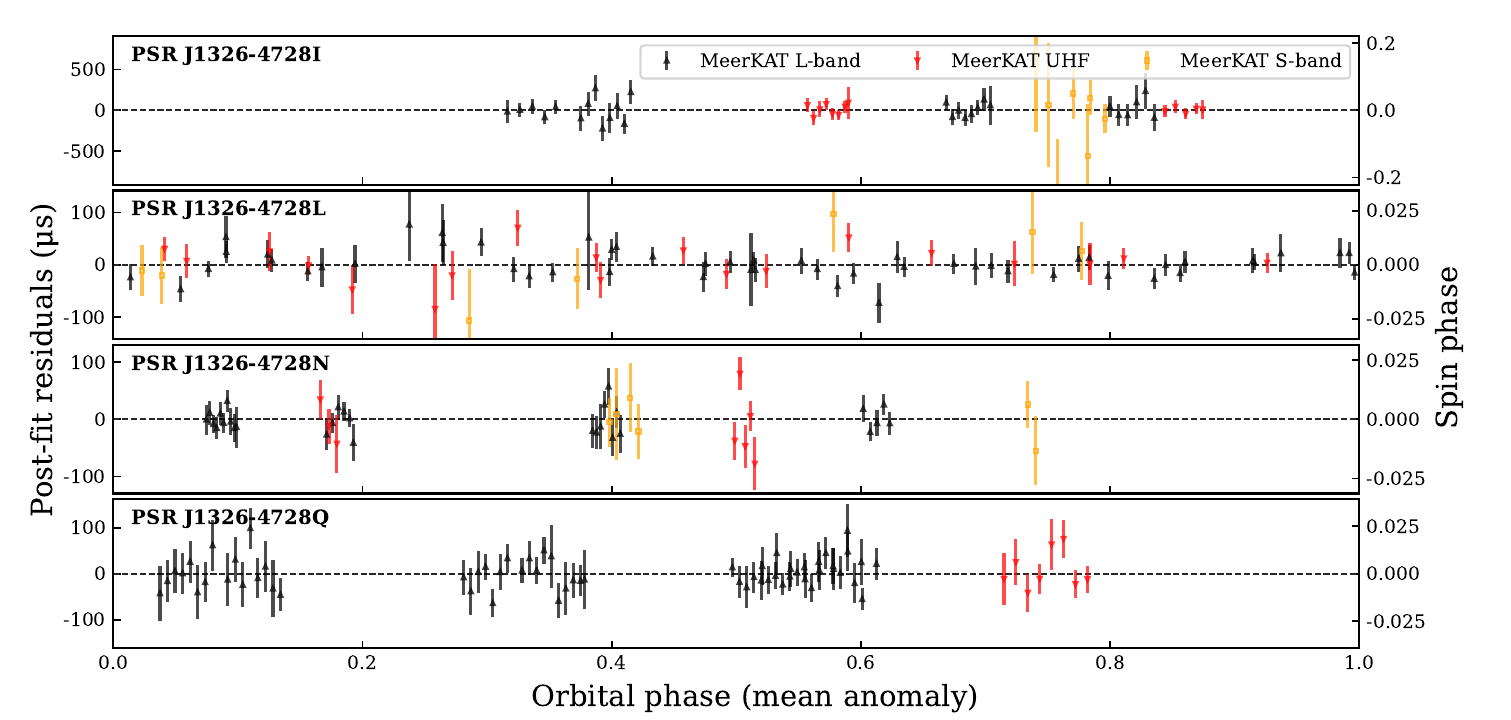}
 \caption{MeerKAT timing residuals for I, L, N, and Q, plotted as a function of orbital phase, showing the orbital coverage for each pulsar. Note that the residuals are relative to each observation (timing jumps applied) and not to a global reference. Included observations comprise all TRAPUM/MeerTIME MeerKAT epochs listed in Table~\ref{observations_table}, except for the S band and 28 July 2025 UHF observations for Q, and including an additional L-band PTUSE 14 February 2024 observation for L.}
 \label{orbital_residuals}
 
\end{figure*}

Timing residuals for A, B, C, D, E, G, H, and K are shown in Fig.~\ref{timing_residuals}. Tables~\ref{timing_isolated_appendix} and \ref{timing_binary_appendix} present the timing solutions and derived parameters for isolated and binary pulsars, respectively. Residuals are predominantly white and centred around zero, with $\chi^2_\mathrm{r}<2.0$ in all cases except for pulsar B ($\chi^2_\mathrm{r}\approx3.7$). Pulsar B showcases significant timing noise associated with its extreme eclipsing behaviour, which introduces flux variability and time-varying ToA delays within individual observations. An upcoming publication dedicated to the eclipses of pulsar B is currently under preparation (Colom i Bernadich et al., in prep.), but in this study we note that its ToA residuals are only correlated at the time scale of minutes, while on the scale of hours they are consistent with extreme white noise.

Our analysis yields significant $\dot P_\mathrm{s}=-\dot\nu/\nu^2$ measurements for all eight pulsars, with significances of $10^{3}$--$10^{4}$\textsigma. For pulsars A--E, the measurements are a factor of ten more precise than those reported by \cite{dai2023omcen}. For G, H, and K, these are first-time measurements. For pulsars A and H, we also find strong evidence in favour of the presence of a positive $\ddot\nu$ term in their residuals, of several $10^{-26}$\,s\,s$^{-2}$ in magnitude. In the case of H, it has a significance of 3\textsigma, making it a fringe detection, while in the case of A it reaches 4.7\textsigma. For all other pulsars we derive 3\textsigma~limits of $|\ddot\nu|\lesssim 10^{-25}$\,s\,s$^{-2}$.

For the timed binary pulsars (B, G, H, and K), we measured black widow-like properties (see \citealt{roberts2013spiders,blanchard2025censusbw}), such as short orbital periods ($P_\mathrm{b}=2$--4\,hours), circular orbits ($e<10^{-3}$), and low companion masses ($M_\mathrm{c}<0.1$\,M$_\odot$, as derived from the mass function). Pulsars B, G and K display eclipses, which is commonly observed in black widows \citep[e.g.][]{fruchter1988eclipsing,abbate2024eclipsing}, and remarkably K shows a highly significant orbital period derivative, $\dot P_\mathrm{b}=6.24(16)\times10^{-12}$\,s\,s$^{-1}$, corresponding to a detection significance of 38.8\textsigma~(see Section~\ref{PBdot_section} for a discussion on its origin). Pulsar B also shows a 7.6\textsigma~detection of $\dot P_\mathrm{b}=(3.8\pm0.5)\times10^{-12}$\,s\,s$^{-1}$, although we note this measurement could be biased by intra-observation timing noises. Pulsars G and H present less significant $\dot P_\mathrm{b}$ constraints of 4.2 and 2.8\textsigma.

\subsection{Orbital solutions: Data analysis}

\begin{figure*}[h!]
\centering
 \includegraphics[bb=00 0 454 340, width=0.53\hsize, clip]{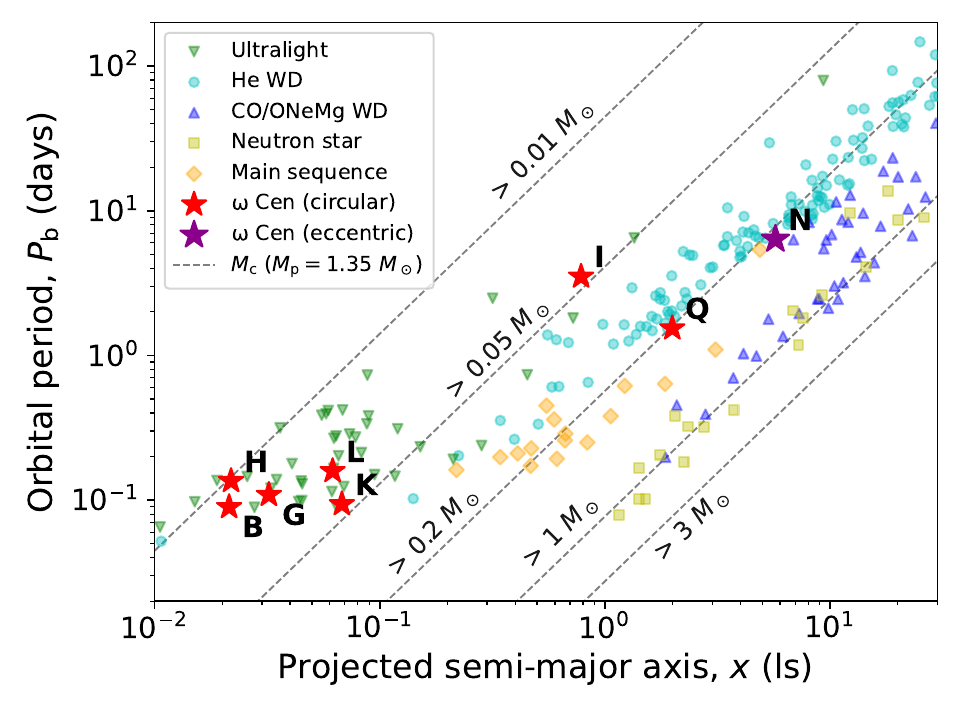}
 \includegraphics[bb=70 0 455 340, width=0.45\hsize, clip]{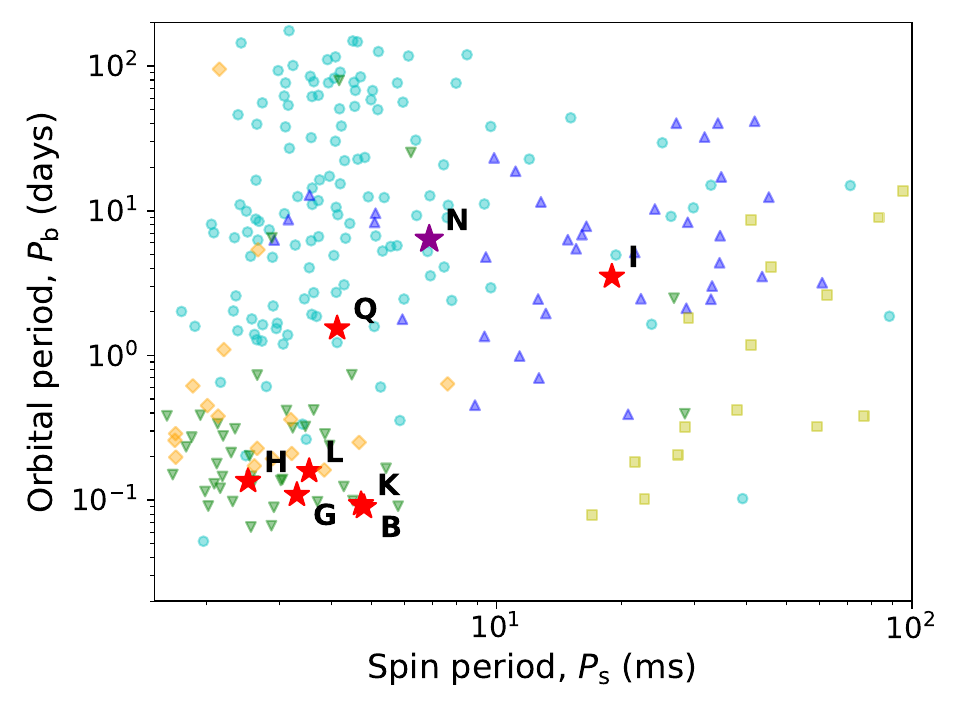}
 \caption{Orbital and spin properties of binary pulsars in \textomega~Cen compared with the Galactic binary populations. The Galactic pulsar parameters and companion nature are taken as listed by the ATNF (\citealt{manchester2005atnf}, version 2.0.1). \textbf{Left:} Orbital period versus projected semi-major axis, with lines indicating constant minimum companion mass. \textbf{Right:} Orbital period versus spin period.}
 \label{binary_populations_figure}
\end{figure*}

For the remaining MSPs in \textomega~Cen, we were unable to derive unique phase-connected timing solutions. In most cases, this was due to the lack of frequent high-S/N detections in Murriyang UWL observations, owing to the intrinsic faintness of the pulsars. Although MeerKAT provided more consistent detections, observing gaps of several months to years prevented us from determining unique timing solutions due to phase ambiguity. Pulsar I was an exception: despite frequent Murriyang UWL detections, we were unable to obtain any viable solution, possibly due to unidentified outliers in the ToAs or an incomplete timing model.

Despite the lack of phase-connected timing solutions, the MeerKAT ToAs enabled us to derive precise orbital parameters for the remaining binary pulsars I, L, N, and Q. For the latter three, MeerKAT provided the only detections, as these pulsars are too faint to be individually detected with Murriyang UWL. For these four pulsars, orbital solutions were obtained using frequency-integrated MeerKAT ToAs. Since the absence of phase connection prevented coherent folding across observing epochs, timing templates were derived from individual high-S/N detections. In the analysis, we included all observations listed in Table~\ref{observations_table}, except for the exceptions listed below.

For Q, the S-band observations and the 28 July 2025 UHF observation were excluded due to non-detections. The absence of S-band detections can be explained by the intrinsic faintness above 2\,GHz. On 28 July 2025, we were unable to recover the pulsar using either direct folding or a targeted \texttt{PRESTO} search. This suggests either a temporary decrease in flux density or signal loss due to the partial data loss reported in Table~\ref{observations_table}. If the non-detection was caused by a temporary flux decrease, eclipsing is unlikely: the observation covered orbital phases 0.41--0.53, well outside superior conjunction at phase 0.25. The non-detection likely resulted from scintillation-induced flux variation combined with partial loss of the observing band, reducing sensitivity to the pulsar signal.

For L, we additionally included the dedicated 14 February 2024 PTUSE MeerTIME L-band observation to improve the orbital solution. As stated in Section~\ref{observations_section}, this observation targeted pulsar L exclusively. Given the lack of Murriyang UWL detections and the limited number of APSUSE observations, including this PTUSE observation significantly improved the constraints on the orbital parameters.

We fit the ToAs with \texttt{TEMPO2}, including only $\nu$ and the Keplerian orbital parameters as free parameters. We additionally fitted arbitrary timing offsets (timing jumps) between observations to enable a global fit despite the lack of phase connection. For pulsars I, L, and Q, we used the ELL1 timing model. Pulsar N was modelled using the Keplerian parameters of the Damour–Deruelle binary model \citep[DD;][]{damour1986general}, including $e$, $\omega$, and the time of periastron passage, $T_0$.

\subsection{Orbital solutions: Results}

In Fig.~\ref{orbital_residuals}, we give the orbital phase coverage of MeerKAT ToAs for I, L, N, and Q. We note that the figure does not display absolute timing residuals because arbitrary timing jumps were fitted independently for each observation. Instead, it shows relative residuals within each observation. The (not phase-connected) orbital solutions and derived parameters are presented in Table~\ref{orbital_binary_appendix}. Figure~\ref{binary_populations_figure} also illustrates the orbital and spin parameters (from Table~\ref{summary_table}) of the \textomega~Cen binaries compared with Galactic field binaries from the ATNF pulsar catalogue\footnote{\url{http://www.atnf.csiro.au/research/pulsar/psrcat}} \citep{manchester2005atnf}.

The orbital parameters of L are consistent with those reported by \cite{chen2023omcen}. The system is black widow-like ($P_\mathrm{b}=3.81$\,hours, $e\lesssim10^{-3}$, $M_\mathrm{c}<0.1$\,M$_\odot$), similar to B, G, H, K, and other MSPs with ultra-light companions (Fig.~\ref{binary_populations_figure}). In addition, it also displays regular eclipses at superior conjunction.

In contrast, the newly derived orbital parameters for I, N, and Q differ significantly from the preliminary values presented by \cite{chen2023omcen}, which were based on only two observations. All three systems exhibit substantially longer orbital periods than previously reported, with $P_\mathrm{b}=3.51$, 6.36, and 1.54\,days for I, N, and Q. Pulsar N also shows a significant eccentricity of $e=0.0928$, reported here for the first time. Both N and Q have $M_\textrm{c}=0.2$--0.3\,M$_\odot$ companions, as inferred from their mass functions. This and their recycled nature are typical characteristics of MSP--He white dwarf binaries (Fig.~\ref{binary_populations_figure}, see Section~\ref{N_section} for a discussion of the eccentricity of N).

Finally, pulsar I presents an unusual configuration of parameters (Fig.~\ref{binary_populations_figure}). It has a circular orbit and a low-mass companion ($M_\mathrm{c}\sim0.05$\,M$_\odot$ from the mass function), similar to those of black widows. However, it has a significantly longer orbital period of $P_\mathrm{b}=3.51$~days, and it is the slowest-spinning pulsar in \textomega~Cen, with $P_\mathrm{s}=18.95$\,ms. It is most similar to PSR J1502$-$6752 in the Galactic field ($P_\mathrm{s}=26.7$\,ms, $P_\mathrm{b}=2.48$\,days, $M_\mathrm{c}\sim0.025$\,M$_\odot$, \citealt{ng2014htru}), which was also notable due to its unusual set of spin and orbital parameters \citep{keith2012htru}.

\section{Searches for high-energy emission counterparts}\label{high_section}

\subsection{Fermi LAT analysis of $\gamma$-rays}

\textomega~Cen is associated with the Fermi-Large Area Telescope \citep[LAT,][]{Atwood09} source 4FGL\,J1326.6$-$4729. Modelling 4FGL\,J1326.6-4729 as a log parabola in 4FGL-DR4 we get $4.3\times10^{-9}$\,photons\,cm$^{-2}$\,s$^{-1}$, or $5.9\times10^{-12}$\,erg\,cm$^{-2}$\,s$^{-1}$ for the integral fluxes in the 0.1--30 GeV band. Because the spectrum is strongly curved, this is identical to the full Fermi band to within less than 1\%. Relative to the LAT PSF, the MSP position offsets from the point source are negligible.

Following the analysis of \cite{dai2023omcen} of pulsars A--E, where no $\gamma$-ray pulsations were found, we folded the $\gamma$-ray photons of G, H, and K to constrain their pulsed flux. We used 17\,yr of P8\_R3\_SOURCE 8 data \citep{Atwood13,Bruel18}, selecting events with reconstructed direction within 3\,deg of the cluster centre and reconstructed energies in the 0.1--30\,GeV range. As for \cite{dai2023omcen}, we considered both the total data span and only the interval constrained by the radio timing data. We used the 4FGL-DR4 sky model \citep{Abdollahi22,Ballet23} to estimate photon weights \citep{Bickel08}, and the probability that a photon is associated with $\gamma$-ray source or background sources. Using the timing solutions, we computed the photon phase \citep{Ray11} and calculated the weighted H-test statistic \citep{Kerr11}.  We found $H<7$ for all three pulsars, indicating no detections.

We determined the maximum amount of the $\gamma$-ray flux from 4FGL\,J1326.6$-$4729 originates from individual pulsars via Monte Carlo simulations. For each pulsar in A--E, G, H, and K, we assumed a typical pulse profile \citep{smith2023third}, $f(\phi)$, composed of two Gaussians with widths $\sigma$ separated by 0.45 in phase, with a flux ratio of 1.5, and a pulsed and un-pulsed components with contributions $w$ and $1-w$. For each experiment, we drew a uniform random variable $u\in[0,1]$ and assigned the photon to the background if $u>w$, otherwise to the source. For the photons assigned to the pulsar, we drew a phase randomly from $f(\phi)$, and we then calculated $H$. We increased $w$ from 0 to 0.2 at a 0.01 step, considering that a pulsar would be detectable when the median H-test exceeds 25. We found a typical $w$ threshold of 10--12\% for $\sigma$ values of 0.02--0.03. This means that, unless the pulse profile is unusually broad, no more than $\sim$10\% of the total $\gamma$-ray flux comes from any of the searched individual pulsars.

\subsection{X-rays}

In Figs.~\ref{chandra_map} and \ref{counterparts_plots} (Appendix~\ref{chandra_appendix}), we compare the pulsar positions with Chandra\footnote{\url{https://science.nasa.gov/mission/chandra/}} X-ray source positions. \citet{dai2023omcen} previously reported that the timing positions of pulsars A, B, C, and E coincide with previously identified Chandra sources \citep{henleywillis2018MNRAS.479.2834H,zhao2022census}. Using our updated timing positions (Table~\ref{timing_binary_appendix}), we additionally found that pulsars G and K are positionally consistent within 1\textsigma~with Chandra sources 24f and 21d from \citet{henleywillis2018MNRAS.479.2834H}, which have positional uncertainties of $\sim$0.4\arcsec. Pulsar H is also marginally consistent ($\sim$3\textsigma~overlap) with source 14c. Finally, based on the localised position reported by \citet{chen2023omcen}, pulsar L may be associated with source 32d at the $\sim$3\textsigma~overlap.

In addition to its X-ray counterpart, pulsar G is also coincident with ATCA radio source 4, whose position was reported by \citet{dai2020omcen}, as shown in Fig.~\ref{counterparts_plots}. The ATCA and X-ray counterparts are themselves coincident with a red giant or sub-giant star identified in Hubble Space Telescope\footnote{\url{https://science.nasa.gov/mission/hubble/}} (HST) observations \citep{cool2013hst/acs}, which could potentially be responsible for the observed X-ray and radio emission.

If the X-ray emission is produced by the pulsars, the most direct way to confirm the associations is through the detection of X-ray pulsations. Motivated by this, we searched for X-ray pulsations and derived corresponding flux limits.

\subsubsection{NICER analysis}

\textomega~Cen was observed 23 times in 2023--2025 with NICER\footnote{\url{https://science.nasa.gov/mission/nicer/}} \citep{gendreau2012nicer}. The main target of these observations was the quiescent LMXB found by \citet{rutledge2002transient}, located on the East side of the cluster. The NICER field of view (FoV) is 30 arcmin$^2$ and contains A, G, H, and K (Fig.~\ref{chandra_map}).

We downloaded all observations and applied the \texttt{nicerl2} pipeline\footnote{\url{https://heasarc.gsfc.nasa.gov/docs/nicer/analysis_threads/nicerl2/}}. The observations were all done after the onset of the light leak issue\footnote{\url{https://heasarc.gsfc.nasa.gov/docs/nicer/analysis_threads/light-leak-overview/}}, and some of them had nominal 0 exposure due to the conservative default conditions of the pipeline on the acceptable undershoot rates. In principle, the light leak should not heavily affect pulsar searches, and therefore, we tried to use more permissive conditions to recover part of the exposure time by tweaking the \texttt{threshfilter}, \texttt{underonlyscr}, \texttt{mpugtiscr}, \texttt{lowmemscr} options. However, we could not improve the available exposure by more than 10\%.
We selected events in the 0.3--12\,keV range.

We folded all NICER data with PINT's \citep{luo2021pint} \texttt{photonphase} script. Before folding, we applied the barycentric correction at the pulsar positions on the cleaned event files. Subsequently, we used the radio solutions derived in this work as references (Tables~\ref{timing_binary_appendix} and \ref{timing_isolated_appendix}), including a TZRMJD value corresponding to a reference radio ToA. We re-fit the radio solutions to use the TDB time scale, the FB90 time ephemeris and the IAU2000B terrestrial-to-celestial transformation, which are supported by PINT, and the DE430 JPL planetary ephemeris.

For each observation, we gave the NICER event list as input, along with the parameter files from the radio solutions. We used the \texttt{--addphase} option to save the pulse phase of each photon to an output file, and we calculated total histograms of pulse phases from all 23 observations. We calculated the H test \citep{deJager1989test} to search for any modulation of the X-ray flux. We repeated the procedure by considering time intervals when CXOU 132619.7–472910.8 was quiescent, and by considering different energy ranges (e.g. 0.3--3\,keV, given the soft spectrum of many non-accreting X-ray pulsars). There was no case where the H test gave a result above 5 for $M=1$ or 2, where $M$ is the number of harmonics that the H test returns. This is well inside the range of values expected for pure noise.

\subsubsection{XMM-Newton analysis}

XMM-Newton observed \textomega~Cen for 40\,ks in 2001 to investigate the faint X-ray sources in the GC \citep{Gendre2003A&A...400..521G}. We aimed to constrain the X-ray emission from pulsar radio positions.

We reduced the observation using \texttt{pysas-1.4.8}\textbf{\footnote{\url{https://github.com/XMMGOF/pysas}}}, a Python wrapper of XMM Scientific Analysis System (XMMSAS). We reprocessed the data using the current calibration files and extracted images from the PN, MOS1, and MOS2 detectors. To quantify the X-ray emission, we demarcated circles of 4\arcsec~around individual pulsar positions and extracted the photon events from these regions. In all cases except for pulsar N, we find that the observed integrated counts are consistent with the typical background level in the cluster of $56\pm7$, sampled from random points with no excess. For N, we observe a contamination from a nearby X-ray source (J132648.725$-$473124.90), also identified in deep Chandra observations \citep[source 33l in][also shown to the right of N in Fig.~\ref{chandra_map}]{henleywillis2018MNRAS.479.2834H}.

The background count rate is $\sim5.5\times10^{-4}$ counts\,s$^{-1}$, which corresponds to a net absorbed flux of $5.3\times10^{-15}$\,ergs\,cm$^{-2}$\,s$^{-1}$. This upper limit is consistent with the estimated fluxes from the Chandra observation \citep{henleywillis2018MNRAS.479.2834H}, which have recorded fainter objects of $\sim10^{-16}$\,ergs\,cm$^{-2}$\,s$^{-1}$ in their survey. We conducted a source detection search and cross-matched the detections with the radio pulsar positions, but we found no counterparts at the pulsar positions.

We note that our measured proper motion magnitudes are of a few milliarcseconds per year, while the point spread function of XMM-Newton and the position uncertainties of detected sources are typically few arcseconds. Thus, the two-decade-long gap between the X-ray observations and the radio observations will not result in significant deviations in the position of the X-ray counterparts.

\section{Dynamical constraints on \textomega~Cen}\label{dynamics_section}

\subsection{Proper motions}

The measured pulsar proper motions (Tables~\ref{timing_isolated_appendix} and \ref{timing_binary_appendix}) are broadly consistent with the known kinematics of the cluster. Their mean values and standard deviations are $\langle\mu_\mathrm{RA},\,\mu_\mathrm{DEC}\rangle = -3.5\pm0.7,\,-7.5\pm1.5$\,mas\,yr$^{-1}$, which coincides with the bulk motion of \textomega~Cen, $\mu_\mathrm{RA},\,\mu_\mathrm{DEC} = -3.1925,\,-6.7445$\,mas\,yr$^{-1}$ \citep{gaia2018gcpm} and the $\sigma_\mathrm{v}\sim0.7$--0.8\,mas\,yr$^{-1}$ velocity dispersion reported by \cite{haberle2025dispersion}. This reinforces the physical nature of our first-time proper motion measurements. Future measurements of increased precision will allow us to probe the internal dynamics of the cluster even further.

\subsection{Acceleration constraints from observed spin-downs}\label{accelerations_section}

To derive limits for the LOS acceleration of individual MSPs due to the internal cluster field, $a_\mathrm{LOS,c}$, we first computed the Shklovskii and relative Galactic acceleration field contributions to $\dot P_\mathrm{s}$ \citep{shklovskii1970possible,damour1992orbital} via
\begin{equation}\label{shk_acc}
\frac{\dot P_\textrm{Sh}}{P_\textrm{s}}=
\frac{1}{c}|\vec{\mu}|^2 d\mathrm{,\,\,\,}
\frac{\dot P_\textrm{G}}{P_\textrm{s}}=
\frac{1}{c}\vec{K}_0\cdot(\vec{a}_\textrm{\textomega~Cen}-\vec{a}_\textrm{SSB})\mathrm{,}
\end{equation}
where $c$ is the speed of light, $\vec{\mu}$ is the proper-motion vector, $\vec{K}_0$ is the unit vector from the solar-system barycentre (SSB) to \textomega~Cen, and $\vec{a}_\textrm{\textomega~Cen}$ and $\vec{a}_\textrm{SSB}$ are the Galactic-field acceleration vectors at the locations of the cluster and the SSB, respectively.

We adopted the values of measured proper motions (Tables~\ref{timing_isolated_appendix} and \ref{timing_binary_appendix}) and the dynamical cluster distance of $d = 5494 \pm 61$\,pc \citep{haberle2025dispersion}. We chose this distance estimate as it is the most precise available in the literature. The resulting Shklovskii contributions span $(1.2$–$5.5)\times10^{-10}$\,m\,s$^{-2}$ across the pulsar sample. The relative Galactic acceleration contribution was computed using the Galactic potential of \citet{mcmillan2017distribution}, assuming a Galactocentric distance of 8.2\,kpc for the SSB and the cluster distance quoted above. This yielded a relative LOS acceleration of $-1.04\times10^{-10}$\,m\,s$^{-2}$ for all pulsars.

We propagated the uncertainties from the proper motion and cluster distance into the Shklovskii term via Eq.~\ref{shk_acc}, resulting in acceleration uncertainties of the order of $10^{-11}$–$10^{-10}$\,m\,s$^{-2}$. We also assigned a conservative 20\% uncertainty to the Galactic acceleration field from \citet{mcmillan2017distribution} (equivalent to $\sim$10$^{-11}$\,m\,s$^{-2}$).

Upper limits on the LOS acceleration due to the internal cluster field for individual MSPs were then obtained by applying these correcting factors to the observed spin-down via
\begin{equation}\label{acceleration_upper_limits}
a_\mathrm{LOS,c}<c\cdot(\dot P_\mathrm{s}-\dot P_\mathrm{Sh}-\dot P_\mathrm{G})/P_\mathrm{s}\mathrm{,}
\end{equation}
where the upper-limit nature arises because the intrinsic spin-down contribution to $\dot P_\mathrm{s}$ cannot be independently quantified. The individual values of $\dot P_\mathrm{Sh}$, $\dot P_\mathrm{G}$, and upper limits on $a_\mathrm{LOS,c}$, are listed for each pulsar in Tables~\ref{timing_isolated_appendix} and \ref{timing_binary_appendix}.

\subsection{Orbital period derivative contributions}\label{PBdot_section}

We investigated the origin of the $\dot{P}_\mathrm{b}$ measured in the timing binaries. We considered whether they might arise from the Shklovskii, Galactic, and cluster contributions discussed in the section above.

Eqs.~\ref{shk_acc} and \ref{acceleration_upper_limits} apply analogously for $P_\mathrm{b}$ and $\dot P_\mathrm{b}$. For pulsar K, which has the most significant measurement ($\dot P_\mathrm{b}=6.24(16)\times10^{-12}$\,s\,s$^{-1}$), we estimate contributions of $\sim7\times10^{-15}$\,s\,s$^{-1}$ and $-3\times10^{-15}$\,s\,s$^{-1}$ from the Shklovskii effect and Galactic acceleration, respectively. These are three orders of magnitude smaller than the observed value, and therefore unable to account for the measurement. Similarly, using the upper limit on the LOS cluster acceleration for K, $a_\mathrm{LOS,c}<(-6.4\pm0.6)\times10^{-10}$\,m\,s$^{-2}$, we predict a cluster contribution to $\dot P_\mathrm{b}$ of $<-2\times10^{-14}$\,s\,s$^{-1}$, which is also significantly smaller and of opposite sign. Similar magnitudes are at play for pulsars B, G, and H, which nonetheless have much less significant measured $\dot P_\mathrm{b}$ values.

This is indicative of the measured $\dot P_\mathrm{b}$ not being dominated by cluster dynamical effects. Instead, they may arise from intrinsic orbital period evolution, as observed in several black widow systems, possibly driven by interactions between the binary and material ablated from the companion \citep[e.g.][]{shaifullah2016blackwidow}. Interaction with a third body is disfavoured given that it would also be detected as a significantly larger $\dot{P}_\mathrm{s}$, which is not observed. Hence, in light of this conclusion, we exclude the $\dot P_\mathrm{b}$ measurements from the dynamical constraints analysis of \textomega~Cen.

\subsection{Second spin frequency derivatives}

We briefly investigated whether the $\ddot\nu$ measurements for pulsars A and H could be interpreted in terms of jerk. The time derivative of the Shklovskii acceleration (Eq.~\ref{shk_acc}) depends on the radial velocity of the pulsars. To estimate their value, we adopt the cluster radial velocity of $235.12$\,km\,s$^{-1}$ reported by \cite{gaia2018gcpm}. The expected jerk is of the order of $10^{-25}$\,m\,s$^{-3}$, corresponding to $\ddot\nu\sim10^{-31}$\,Hz\,s$^{-2}$ for an MSP with $\nu=200$\,Hz, far below the measured values. Similar jerk magnitudes are expected for objects orbiting in Galactic field/halo, while values as large as $10^{-20}$\,m\,s$^{-3}$, corresponding to $\ddot\nu\sim10^{-27}$\,Hz\,s$^{-2}$ for a $\nu=200$\,Hz MSP, can occur for pulsars in GCs \citep[e.g.][]{freire2017tuc47}.

The measured $\ddot\nu\sim10^{-26}$--$10^{-25}$\,Hz\,s$^{-2}$ values for A and H are substantially larger than the contributions discussed above. Instead, they are comparable to $\ddot\nu$ values observed in MSPs in wide-orbit hierarchical triple systems or undergoing stellar encounters \citep[e.g.][]{grunthal2024triple,dutta2025three}. However, the $\ddot\nu$ values measured in our study have low statistical significance and may instead be caused by unmodelled timing noise. We therefore refrain from assigning them a physical interpretation and defer any conclusions until the timing precision improves. For this reason, for the dynamical analysis of \textomega~Cen, we ignored the $\ddot\nu$ measurements.

\subsection{The mass profile of \textomega~Cen}

As is commonly done in GCs \citep[e.g.][]{prager2017timingter5,freire2017tuc47,dai2023omcen}, we used the $a_\mathrm{LOS,c}$ constraints to assess their compatibility with King model density and potential \citep{king1962profile}. Following \cite{fhn+05} and \citet{dai2023omcen}, we adopted a King model for the mass distribution, which yields a radial acceleration field via
\begin{equation}\label{cluster_acceleration}
a_\mathrm{c}(r^\prime)=\frac{9\sigma_\mathrm{v}^2 d}{r_\mathrm{c}}\frac{1}{r{^\prime}^2}
\left(\frac{r^\prime}{\sqrt{1+{r^\prime}^2}}-\sinh^{-1}r^\prime\right)\mathrm{,}
\end{equation}
where $\sigma_\mathrm{v}$ is the central velocity dispersion, $r_\mathrm{c}$ is the core radius, $r^\prime=r/r_\mathrm{c}$ is the distance from the cluster centre normalised by the core radius, and $d$ is the cluster distance, which we took to be $d=5494$\,pc from the dynamical measurement of \citet{haberle2025dispersion}. From this, and assuming $r_\mathrm{c}=2.88$\arcmin~ \citep{baumgardt2018catalogueGCs} and $\sigma_\mathrm{v}=0.81$\,mas\,yr$^{-1}$ \citep{haberle2025dispersion}, we derived the maximum and minimum expected values of $a_\mathrm{LOS,c}$ as a function of projected angular separation from the cluster centre, $\theta$.

\begin{figure}[t!]
\centering
 \includegraphics[bb=00 0 455 335, width=\hsize, clip]{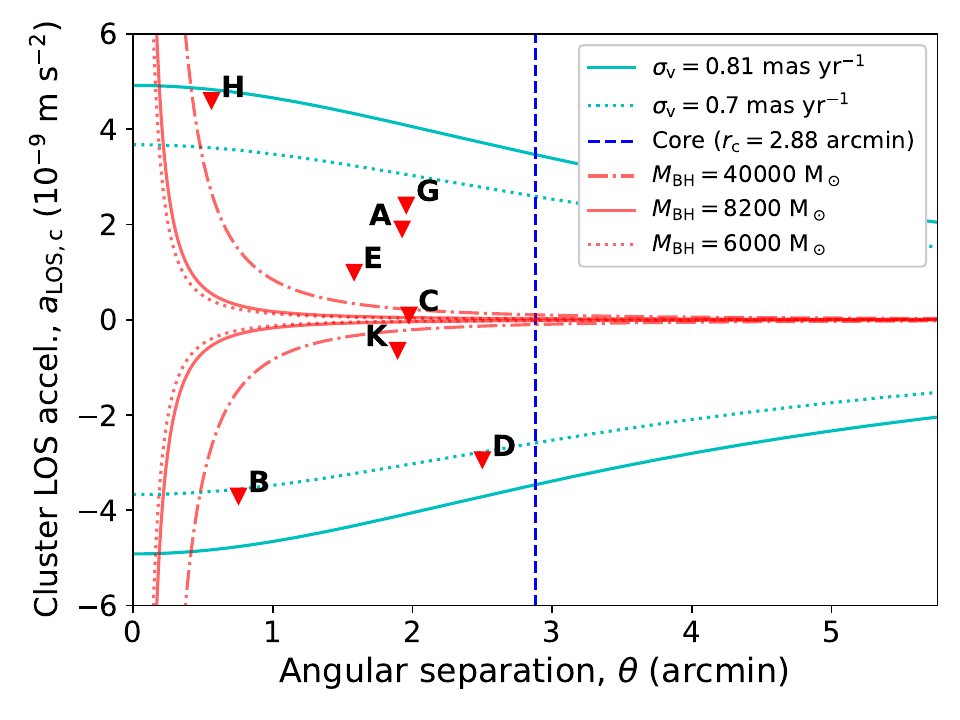}
 \caption{Upper limits on LOS accelerations for pulsars against angular separation from the cluster centre. The plot also shows the maximum and minimum LOS accelerations expected from the King potential, assuming $\sigma_\mathrm{v}$ and $r_\mathrm{c}$ from \citet{baumgardt2018catalogueGCs} and \citet{haberle2025dispersion}, as well as the contributions from central compact masses of 6000, 8200, and $4\times10^{4}$\,M$_\odot$.}
 \label{acceleration_constraints}
\end{figure}

In Fig.~\ref{acceleration_constraints}, we compare the measured $a_\mathrm{LOS,c}$ upper limits with their predicted maximum values as a function of projected angular separation from the cluster centre. For completeness, we also added the limits imposed by the King model assuming $\sigma_\mathrm{v}=0.70$\,mas\,yr$^{-1}$ \citep{baumgardt2018catalogueGCs}, as done by \citealt{dai2023omcen}, as well as the maximum and minimum contributions from a central compact object or dark mass.

We find that pulsars in \textomega~Cen with measured $\dot{P}\mathrm{s}$ are consistent with the updated King-model cluster mass distribution. Assuming a velocity dispersion of $\sigma_\mathrm{v}=0.81$\,mas\,yr$^{-1}$ as measured by \citet{haberle2025dispersion}, the upper limits on $a_\mathrm{LOS,c}$ are smaller in magnitude than the maximum magnitudes predicted by the model for all pulsars. However, adopting $\sigma_\mathrm{v}=0.7$\,mas\,yr$^{-1}$ as reported by \citet{baumgardt2018catalogueGCs}, pulsars H, B, and D become potential outliers, with upper limits on $a_\mathrm{LOS,c}$ exceeding the model constraints. Overall, our pulsar measurements are more consistent with recent dynamical descriptions of \textomega~Cen from \cite{haberle2025dispersion}

\subsection{Constraints on the central IMBH} 

\citet{haberle2024inmbh} present evidence for a central compact object of at least $8200$\,M$_\odot$ based on stellar proper motions within the central parsec. In contrast, \citet{banares2025omcenconstraints} favour an extended central mass and place an upper IMBH limit of $6000$\,M$_\odot$ from multiple astrometric measurements, including the timing measurements from \citet{dai2023omcen}. In this section, we constrain the maximum magnitude of a central dark mass according to the observed pulsar LOS accelerations.

The pulsars analysed in our work lie outside the region of influence of an IMBH with mass $10^3$--$10^4$\,M$_\odot$.  Figure~\ref{acceleration_constraints} illustrates the maximum and minimum LOS acceleration contributions expected from central point masses of $8200$ and $6000$\,M$_\odot$. Even for pulsars H and B, the closest to the cluster centre, at angular separations of $\theta=0.56$\arcmin~and 0.76\arcmin, the acceleration induced by an IMBH in this mass range would be marginal compared to the broader cluster contribution. H and B would, however, be sensitive to an IMBH larger than $10^4$\,M$_\odot$, as exemplified in Fig.~\ref{acceleration_constraints} by the maximum contribution from a central mass of $4\times10^4$\,M$_\odot$. To quantify this, we performed a likelihood analysis for black hole masses from $0$ to $10^6$\,M$_\odot$, expressed as
\begin{equation}\mathcal{L}(M_\mathrm{BH})=\prod_\mathrm{PSR}p(a_\mathrm{LOS,c},\theta\,|M_\mathrm{BH})\mathrm{,}\end{equation}
where $p(a_\mathrm{LOS,c}, \theta \mid M_\mathrm{BH})$ is the probability of a pulsar exhibiting a LOS acceleration $a_\mathrm{LOS,c}$ at an angular separation of $\theta$, assuming the King density profile,
\begin{equation}\rho(r^\prime)=\frac{\rho_0}{(1+{r^\prime}^2)^{3/2}}\mathrm{,}\end{equation}
where $\rho_0$ is the central density, and the cluster gravitational field from equation~\ref{cluster_acceleration} ($\sigma_\mathrm{v}=0.81$\,mas\,yr$^{-1}$, $r_\mathrm{c}=2.88$\arcmin), with the addition of a contribution from a central mass $M_\mathrm{BH}$,
\begin{equation}a_\mathrm{BH}(r)=-G\frac{M_\mathrm{BH}}{r^2}\mathrm{.}\end{equation}

We also estimated a distribution for the intrinsic spin-down contribution. The surface magnetic field strengths ($B_\mathrm{s}$) of GC MSPs follow a log-normal distribution with $\mu_{\log B}=8.47$ and $\sigma_{\log B}=0.33$ \citep{prager2017timingter5}. Assuming a braking index $n=3$, we converted this into a distribution for the intrinsic spin-down ($\dot{P}_\mathrm{s,int}$) contribution to the observed LOS acceleration,
\begin{equation}a_\mathrm{int}=c\left(\frac{\dot{P}_\mathrm{s,int}}{P_\mathrm{s}}\right)=7.96\times10^{-10}\left(\frac{B_\mathrm{s}}{2\times10^{8}\,\mathrm{G}}\right)^2\left(\frac{2\,\mathrm{ms}}{P_\mathrm{s}}\right)^2\,\mathrm{m\,s}^{-2}\mathrm{,}\end{equation}
which is subtracted from the observed $\dot P_\mathrm{s}$ along with $ \dot P_\mathrm{Sh}$ and $\dot P_\mathrm{G}$ in equation~\ref{acceleration_upper_limits} to obtain a `true' $a_\mathrm{LOS,c}$ distribution. In addition, we also accounted for a Gaussian distribution in the measured $a_\mathrm{LOS,c}$ upper limits given by the 1\textsigma~uncertainties in the Shklovskii and relative Galactic acceleration terms (see Section~\ref{accelerations_section}). The uncertainties and the contribution of $B_\mathrm{s}$ were accounted for via a Monte Carlo approach, where we sampled millions of possible `true' $a_\mathrm{LOS,c}$ values and computed the likelihood averaged over these realisations for each $M_\mathrm{BH}$ value.

\begin{figure}[t!]
\centering
 \includegraphics[bb=00 0 455 335, width=\hsize, clip]{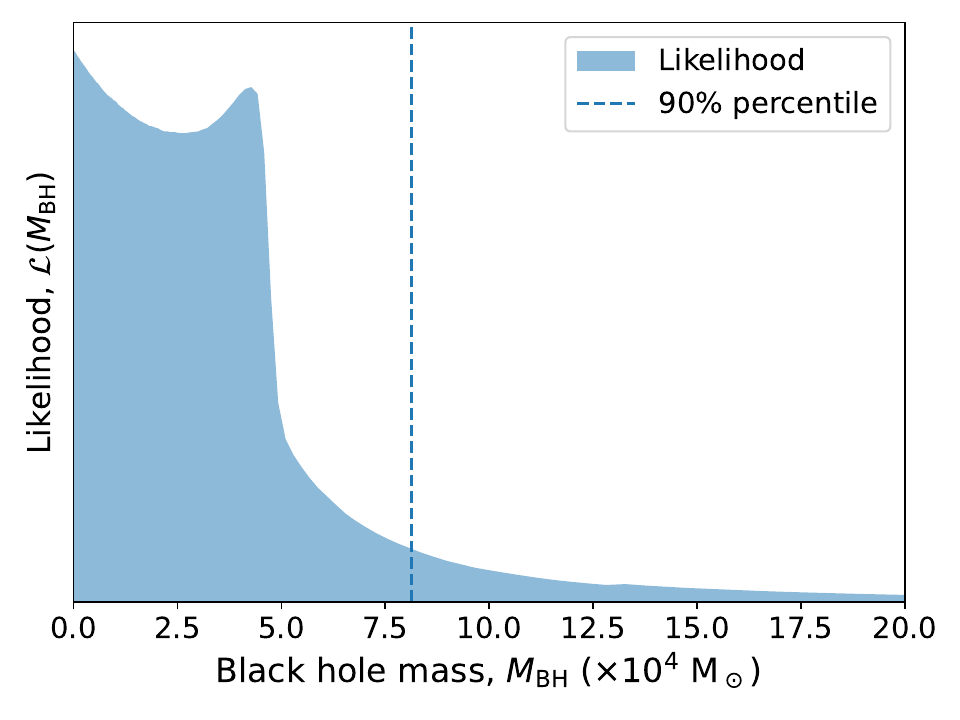}
 \caption{Likelihood distribution for a central black hole or dark mass $M_\mathrm{BH}$, derived from the $\dot P_\mathrm{s}$ measurements and LOS acceleration constraints shown in Fig.~\ref{acceleration_constraints}.}
 \label{mass_constraints}
\end{figure}

The likelihood is shown in Fig.~\ref{mass_constraints}, with the 90\% confidence interval highlighted. The distribution shows elevated likelihoods for $M_\mathrm{BH} \lesssim 4\times10^4$\,M$_\odot$, with a primary peak at $M_\mathrm{BH}=0$\,M$_\odot$ and a secondary peak near $M_\mathrm{BH}\approx 4\times10^4$\,M$_\odot$, followed by a rapid decline at higher masses. From the 90\% percentile of the distribution, we derive an upper limit of $M_\mathrm{BH}<8\times10^4$\,M$_\odot$ (90\% confidence). We note that this analysis constrains not only the mass of a central compact object, but also the total mass of any excess central mass distribution enclosed within the innermost pulsar considered here (pulsar H).

\begin{figure*}[h!]
\centering
 \includegraphics[width=0.49\hsize]{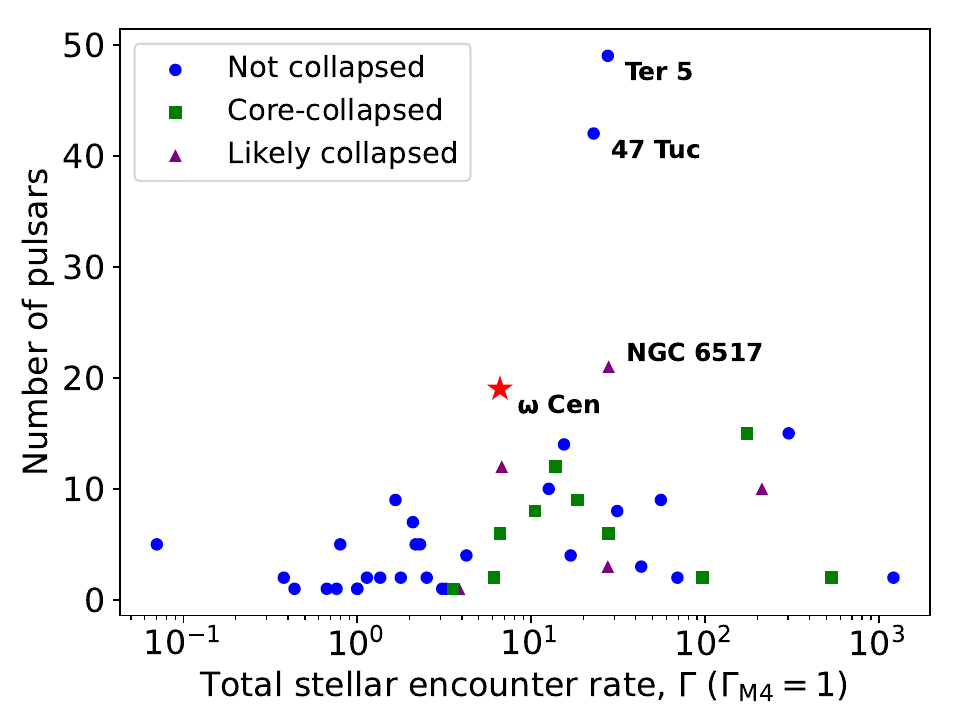}
 \includegraphics[width=0.49\hsize]{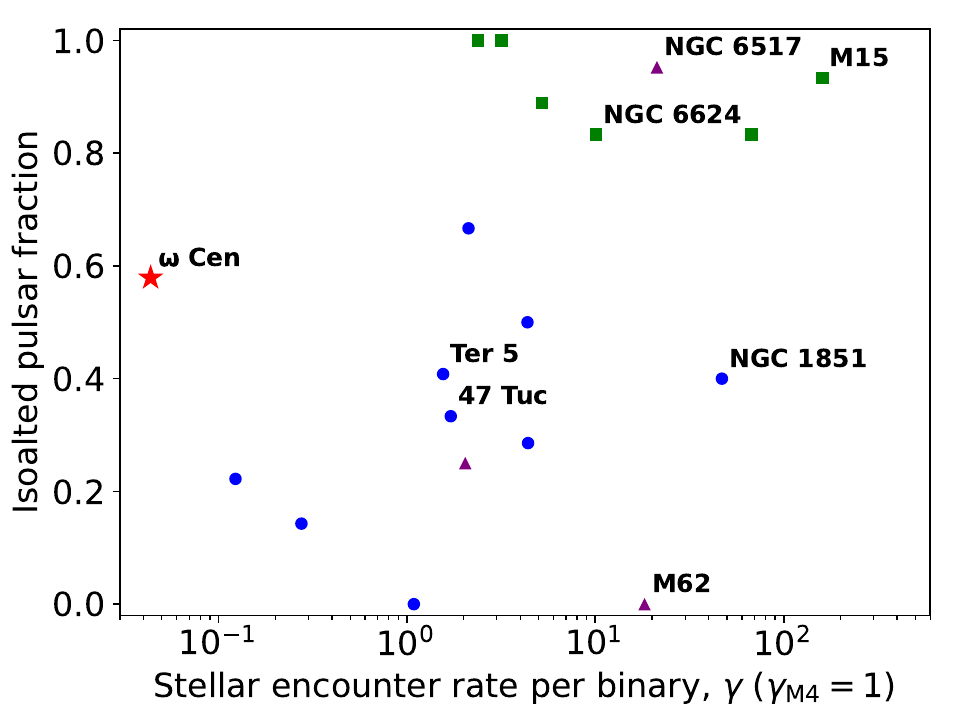}
 \caption{Pulsar populations in globular clusters as a function of stellar encounter rate. \textbf{Left:} Total number of pulsars versus the global encounter rate, with well-studied clusters highlighted. \textbf{Right:} Fraction of isolated pulsars versus the encounter rate per binary for clusters hosting at least five pulsars. Pulsar data are taken from \url{https://www3.mpifr-bonn.mpg.de/staff/pfreire/GCpsr.html}, cluster parameters from \citet{baumgardt2018catalogueGCs}, and core-collapse classifications from \citet{harris2010catalog}. NGC~6517 is treated as core collapsed in several studies \citep[e.g.][]{verbunt2014disruption,yin2024ngc6517}, despite not being classified as such by \citet{harris2010catalog}.}
 \label{encoutner_rates}
\end{figure*}

\section{The MSP populations of \textomega~Cen and their origin}\label{population_section}

\subsection{The encounter rates of \textomega~Cen}

As discussed in the introduction, the total pulsar populations and binary pulsar fraction in GCs are expected to correlate with the total stellar encounter rate, $\Gamma$, and the encounter rate per individual binary, $\gamma$. In this section, we discuss the pulsar population of \textomega~Cen in these terms.

In particular, $\Gamma$ is expected to be a predictor of the overall MSP content \citep{hui2010ecounters,bahramian2013encounters}. If pulsar recycling is driven by the formation of close binaries in encounters, the MSP population should scale as $\Gamma \propto \rho_\mathrm{c}^{3/2} r_\mathrm{c}^2$, where $\rho_\mathrm{c}$ is the core density \citep{verbunt1987origin}. For \textomega~Cen, we adopt $\rho_\mathrm{c}=1.7\times10^{3}$\,M$_\odot$\,pc$^{-3}$ and $r_\mathrm{c}=4.30$\,pc \citep{baumgardt2018catalogueGCs}. Normalising to M4 (NGC~6121, $\rho_\mathrm{c}=9.8\times10^{3}$\,M$_\odot$\,pc$^{-3}$, $r_\mathrm{c}=0.45$\,pc, $\Gamma_\mathrm{M4}\equiv1$), we obtain $\Gamma_\textrm{\textomega~Cen}=6.6$.

In contrast, $\gamma$ quantifies the likelihood that an existing binary will undergo further interactions, producing eccentric systems, exchanges, or complete disruption. In general, larger $\gamma$ values would predict a larger fraction of isolated pulsars, and it is expected to scale as $\gamma \propto \rho_\mathrm{c}^{1/2} r_\mathrm{c}^{-1}$ \citep{verbunt2014disruption}. Normalising to M4 yields $\gamma_\textrm{\textomega~Cen}=0.044$.

For comparison, Figure~\ref{encoutner_rates} plots the pulsar populations and encounter rates of \textomega~Cen against those of other globular clusters. The encounter rates have been computed from the parameters reported in \citet{baumgardt2018catalogueGCs}. These plots show that $\Gamma_\textrm{\textomega~Cen}=6.6$ is a moderate value that arises from the large size of \textomega~Cen despite its low density. On the other hand $\gamma_\textrm{\textomega~Cen}=0.044$ is a remarkably low value when compared to other GCs, reflecting the low density of \textomega~Cen. This would predict a moderately large MSP population composed of undisrupted binaries.

\subsection{The total observed MSP population}

With 19 confirmed MSPs, the population of \textomega~Cen in relation to its $\Gamma$ is consistent with what has been observed in other well-studied clusters. The left plot of Fig.~\ref{encoutner_rates} compares its total observed MSP content and $\Gamma_\textrm{\textomega~Cen}$ with those of other GCs. The known MSP population of \textomega~Cen is only slightly smaller than that of NGC~6517, and slightly less than half of 47~Tuc and Ter~5, which have been explored to large survey depth \citep{yin2024ngc6517,freire2017tuc47,chen2026_47Tuc,padmanabh2024ter5} and have $\Gamma=20$--30 (3--5 times larger than \textomega~Cen). Therefore, the observed number of MSPs is within the expectations.

The size of the observed MSP population of \textomega~Cen is also consistent with its measured $\gamma$-ray luminosity as it was reported in \citet{deMenezes2019gammaray}. From the $L_\gamma = (3.6\pm0.3)\times10^{34}$\,erg\,s$^{-1}$ $\gamma$-ray emission of \textomega~Cen, and assuming an average MSP spin-down power of $(1.8\pm0.7)\times10^{34}$\,erg\,s$^{-1}$ and a $\gamma$-ray efficiency of 0.08, \citet{deMenezes2019gammaray} predicts a total MSP population of 10 to 30 for \textomega~Cen, consistent with the currently observed number of 19. However, accounting for a radio beaming fraction of $\sim$30\% would imply a true total MSP population (observable plus not observable) as large as $\sim$60, which would be significantly higher than the \citet{deMenezes2019gammaray} prediction.

\subsection{The isolated MSP fraction}

The large fraction of isolated MSPs is, however, unexpected, as first highlighted by \citet{chen2023omcen}. Of the 19 confirmed MSPs in \textomega~Cen, 11 are isolated (58\%), despite the cluster's very low $\gamma$ encounter rate. This is seemingly at odds with the relation proposed by \citet{verbunt2014disruption}, which would imply that very few binaries should have been disrupted.

However, it remains unclear how strong or predictive this relation is. The right-hand panel of Fig.~\ref{encoutner_rates} shows the fraction of isolated MSPs in clusters hosting at least five pulsars as a function of their $\gamma$ values. In general, core-collapsed GCs with extreme $\gamma$ values, such as NGC 6624, NGC 6517, and M15, are almost entirely dominated by isolated pulsars \citep{yin2024ngc6517,abbate2022ngc6624,wu2024m15}, whereas non-core-collapsed clusters exhibit a wide range of isolated fractions. In this comparison, \textomega~Cen stands out as the GC with the lowest $\gamma$, yet it has one of the highest isolated MSP fractions among non-core-collapsed GCs. An opposite example is M62, which has a $\gamma$ value comparable to those of core-collapsed clusters, yet whose known MSP population consists entirely of binaries \citep[Fig.~\ref{encoutner_rates};][]{vleeschower2024m62}. These discrepancies suggest that present-day encounter parameters alone are not strong predictors of GC MSP population demographics, and that additional factors likely contribute to shaping their binary fractions.

\subsection{On observational biases favouring isolated discoveries}

Our search strategies (here and in \citealt{chen2023omcen}) favour the detection of isolated pulsars and wide binaries ($P_\mathrm{b}>1$\,day) over compact binary systems. The latter are only detectable as highly accelerated or jerked signals in short observational segments, whereas isolated or mildly accelerated pulsars can accumulate their signals over several hours. In this section, we estimate whether this bias is relevant to the discussion from the previous section.

To estimate the completeness of our searches, we considered the pulsars discovered in \textomega~Cen. Wide binaries such as I, N, and Q were recovered in blind searches using the full integrations, suggesting that the surveys are close to complete for wide binaries above the specific sensitivities ($\sim $10 \textmu Jy at four hours with the L band, and $\sim$20 \textmu Jy at 3 hours with UHF). In contrast, compact binaries with orbital periods of 2--4 hr (B, G, H, K, and L) were detected only in segmented searches. We therefore focused on this population when estimating incompleteness.

Assuming a flux distribution of $\mathrm{d}N/\mathrm{d}S \propto S^{-2}$ for GC MSPs \citep{mcconnell2004luminosity}, we estimated the fraction of compact binaries detectable in our searches. Pulsars B, G, H, K, and L were recovered in 1-hour integrations through the use of jerk searches. Relative to the full integrations, this increases the effective detection thresholds to $\sim$20 \textmu Jy at the L band and $\sim$35 \textmu Jy at UHF. We then compared the number of pulsars that are above the initial and updated thresholds. Under the assumed flux distribution, these thresholds imply that approximately 50\% of compact binaries remain detectable.

This suggests that the underlying compact binary population of \textomega~Cen could contain up to $\sim$10 systems, rather than the five currently detected, raising the total binary number to $\sim$13. The corresponding isolated MSP fraction would decrease from 58\% to 46\%. While this correction is significant, the isolated fraction would still remain unusually high. Moreover, pulsar searches in other GCs are subject to similar selection biases, which would affect all GCs displayed in Fig.~\ref{eccentric_comparison} (albeit to a varying degree depending on survey strategy), meaning that the comparisons drawn here are still meaningful.

\subsection{An over-abundant black widow population}\label{black_widow_section}

Pulsars B, G, H, K, and L have the characteristics of black widows. Five black widows out of eight binaries (63\%) is the largest observed ratio amongst clusters with known sizeable black widow populations. For instance, 47 Tuc, Ter 5, and M28 have black widow/binary fractions of 25\%, 10\%, and 56\% \citep{freire2017tuc47,chen2026_47Tuc,padmanabh2024ter5,douglas2022m28}. Because of the selection biases discussed above, all of these are lower limits.

Globular clusters are known to be a cradle of black widows due to their large encounter rates. \citet{king2003promiscuity} proposed that black widow excesses in GCs (compared to the Galactic plane) arise from secondary stellar encounters, in which the original low-mass companion that recycled the MSP is replaced by a more massive turn-off star. Subsequent stellar encounters may harden the resulting binary even further. Afterwards, Roche-lobe overflow and envelope expulsion through stellar winds shrink and circularise the orbit via tidal interactions. Alternatively, close encounters may have instead hardened neutron star--main sequence progenitor binaries \citep{heggie1975binary,hills1975encounters}.

However, these formation channels are uncertain in \textomega~Cen. The timescale for a binary to experience a stellar encounter with an encounter distance of $r$ can be estimated following \citep{king2003promiscuity} as
\begin{equation}\tau\approx7\times10^{10}\mathrm{\,yr\,}\left(\frac{10^{5}\mathrm{\,pc}^{-3}}{n}\right)\left(\frac{\sigma_\mathrm{v}}{10\mathrm{\,km\,s}^{-1}}\right)\left(\frac{R_\odot}{r}\right)\left(\frac{M_\odot}{m}\right)\mathrm{,}\end{equation}
where $n=3.7\times10^{3}$\,pc$^{-3}$ is the stellar density, estimated from the core mass density of $1.7\times10^{3}$\,M$_\odot$\,pc$^{-3}$ \citep{baumgardt2018catalogueGCs} and a mean stellar mass of $M_\ast=0.45$\,M$_\odot$ derived from the stellar mass function of \textomega~Cen \citep{sollima2007function}, $\sigma_\mathrm{v}\approx20$\,km\,s$^{-1}$ is the core velocity dispersion as measured in \cite{haberle2025dispersion}, and $m=M_\mathrm{p}+M_\mathrm{c}+M_\ast$ is the combined binary and interloper mass. Assuming a typical pulsar binary with $P_\mathrm{b}=1$--10\,days, $M_\mathrm{c}=0.25$\,M$_\odot$, and an encounter distance comparable to the orbital separation (r=$1$--10\,ls), we obtained encounter timescales of $\tau=10^{2}$--$10^{3}$\,Gyr, far exceeding the cluster's age. Furthermore, likely multiple encounters are needed to process a binary into a black widow.

Unless \textomega~Cen has a budget of several thousands of neutron stars in binaries that could potentially be turned into black widows (a requirement difficult to reconcile with estimates of neutron star retention and binary fractions), it is likely that the five observed black widows formed via a channel alternative to encounter rates. Several studies have shown that black widows can form through the evolution of low-mass X-ray binaries or other compact systems without stellar encounters \citep{chen2013formation,benvenuto2015formation,guo2022formation}. This coincides with the observation that \textomega~Cen also has a surprising excess of X-ray binaries \citep{deMenezes2023dynamics}.

Alternatively, the black widow population could have formed in a denser environment than the one observed today. Reducing $\tau$ below the Hubble time ($\tau\lesssim10$\,Gyr) would require increased densities of $10^{4}$--$10^{5}$\,M$_\odot$\,pc$^{-3}$. If such densities existed at some point in the cluster's past, then the formation of this population via encounters becomes plausible.

\subsection{Possible link between black widows and isolated MSPs}

\begin{table*}[t!]
\caption[]{\label{eccentric_comparison}Pulsar and cluster parameters for mildly eccentric MSPs with WD orbital companions in low-density GCs.}
\centering
\begin{tabular}{lcccccccccc}
\hline
\hline \\[-1.5ex]
PSR & $P_\mathrm{s}$ & $P_\mathrm{b}$ & $e$ & $M_\mathrm{c,med}$ & $n$ & $\sigma_\mathrm{v}$ & $\Gamma$ & $\gamma$ & $e_\mathrm{max}$ & Reference \\
 & ms & days &  & M$_\odot$ & pc$^{-3}$ & km\,s$^{-1}$ & $\Gamma_\mathrm{M4}$ & $\gamma_\mathrm{M4}$ &  & \\
\hline \\[-1.5ex]
\textomega~Cen N & 6.88 & 6.36 & 0.093 & 0.27 & $3.7\times10^3$ & 21.1 & 6.61 & 0.044 & $2\times10^{-5}$ & this work \\
M13 I & 3.30 & 18.2 & 0.064 & 0.54 & $4.9\times10^3$ & 8.9 & 1.66 & 0.123 & $2\times10^{-3}$ & \citep{li2025ffaM13} \\
M5 B & 7.95 & 6.86 & 0.138 & 0.13 & $10.3\times10^{3}$ & 7.9 & 2.09 & 0.276 & $4\times10^{-3}$ & \citep{zhang2023timing} \\
NGC 6539 A & 23.1 & 2.62 & 0.212 & 0.35 & $3.65\times10^{3}$ & 6.0 & 0.67 & 0.133 & $10^{-4}$ & \citep{hobbs2004timing} \\
\hline
\hline
\end{tabular}

\tablefoot{Only pulsars with $e\lesssim0.2$ in low-density clusters are included. The pulsar parameters are taken from the ATNF catalogue, while cluster parameters are taken from \cite{baumgardt2018catalogueGCs} (for \textomega~Cen, \citealt{haberle2025dispersion}), with the core mass density converted into stellar density by using an average stellar mass of $M_\ast=0.45$\,M$_\odot$. $e_\mathrm{max}$ estimated via equation~\ref{eccentricity_equation} assuming $t_\mathrm{age}\sim9$\,Gyr, which is a generous value.}

\end{table*}

The formation of MSPs from the evolution of black widow systems has been discussed in the literature. One speculative formation channel is the complete ablation of the companion in black widow binaries \citep{vandenheuvel1988fate}, although this scenario remains poorly constrained.

A recently proposed alternative explains the isolated MSP population of \textomega~Cen via the efficient disruption of binaries owing to the low mass of their companions. \citet{deMenezes2026isolated} shows that if one assumes that isolated MSPs form from disrupted binaries with $M_\mathrm{c}=0.01$\,M$_\odot$, the observed isolated-pulsar fractions across GCs follow a $\propto a_\mathrm{H}^{-1}$ scaling, where $a_\mathrm{H}$ is the ionization threshold calculated from the cluster velocity dispersion and an incoming-star mass of $M_\ast=0.4$\,M$_\odot$. \textomega~Cen in particular fits well within this trend (see Fig.~1 in \citealt{deMenezes2026isolated}), providing a natural explanation for the isolated MSP fraction if these pulsars indeed formed with $M_\mathrm{c}=0.01$\,M$_\odot$ companions.

However, while this is a plausible channel, the \citet{deMenezes2026isolated} scenario requires a sizeable underlying population of black widow-like systems with $M_\mathrm{c}=0.01$\,M$_\odot$ companions. As discussed in Section~\ref{black_widow_section}, the present-day stellar density of \textomega~Cen appears insufficient to sustain such an elevated black widow formation rate. Furthermore, the cases of pulsars Q, and N show that many MSPs also tend to form with more massive companions, as is expected from binary MSP recycling scenarios in low-mass X-ray binaries with orbital periods larger than one day \citep{tauris1999hewd}. This puzzle correlates with the large number of X-ray sources observed in \textomega~Cen \citep{henleywillis2018MNRAS.479.2834H}, which is also abnormally high in relation to its encounter rate and in comparison to other GCs \citep{deMenezes2023dynamics}.

\subsection{The eccentricity of PSR J1326$-$4728N}\label{N_section}

Pulsar N has the widest orbit ($P_\mathrm{b}=6.36$\,days) and a moderate but significant eccentricity ($e=0.093$). With a median mass of $M_\mathrm{c}=0.27$\,M$_\odot$ (from the mass function) and a spin period of $P_\mathrm{s}=6.88$\,ms, the companion is likely a He white dwarf or a low-mass CO white dwarf. The eccentricity is unlikely to have arisen from an exchange, as they are expected to produce much larger eccentricities ($e>0.1$). Owing to the large positional uncertainty of N from the lack of a timing solution, we are unable to search for an optical counterpart arising from a potential main-sequence nature for the companion, although this scenario is disfavoured.

Assuming N originated as a circular binary, as is expected for low-mass MSP--He white dwarf binaries \citep{tauris1999hewd}, we examine whether its eccentricity can be reconciled with the encounter rate in \textomega~Cen. \citet{raiso1995eccentricities} predict that the eccentricities of MSPs in white dwarf binaries in GCs grow over time due to stellar fly-bys, as a function of stellar density, velocity dispersions, and orbital parameters. The maximum obtainable eccentricities \citep{raiso1995eccentricities} are expressed as
\begin{equation}\label{eccentricity_equation}e_\mathrm{max}\approx\mathrm{Max}\Biggl\{\left(\frac{\eta}{400}\right)^{5/2}\left(\frac{P_\mathrm{b}}{\mathrm{day}}\right)^{5/3}
\textrm{, }4\exp\Biggl[-\left(\frac{\eta}{200}\right)^{-3/2}\left(\frac{P_\mathrm{b}}{\mathrm{day}}\right)^{-1}\Biggr]\Biggr\}\mathrm{,}\end{equation}
where
\begin{equation}\eta=\left(\frac{t_\mathrm{age}}{10^9\mathrm{\,yr}}\right)\left(\frac{n}{10^{4}\mathrm{\,pc}^{-3}}\right)\left(\frac{10\mathrm{\,km\,s}^{-1}}{\sigma_\mathrm{v}}\right)\mathrm{,}\end{equation}
Assuming an age of $t_\mathrm{age}\sim9$\,Gyr, we computed $e_\mathrm{max}$ for pulsar N in \textomega~Cen, and a few further MSPs with mild eccentricities and white dwarf companions in low-density globular clusters.

The pulsar, cluster, and maximum predicted eccentricities are listed in Table~\ref{eccentric_comparison}. For pulsar N in \textomega~Cen, we find a maximum expected eccentricity of $2\times10^{-5}$, more than three orders of magnitude smaller than the observed value. Similar discrepancies are seen for other GC MSPs with similar orbital parameters and environments. Pulsar N in \textomega~Cen is another example where the currently observed encounter rates fail to reproduce the observed properties. From equation~\ref{eccentricity_equation}, reproducing the eccentricity of N would require a density of at least several $10^{4}$\,M$_\odot$\,pc$^{-3}$, an order of magnitude larger than the currently observed value.

Alternatively, the eccentricity of N may be explained via another mechanism. For instance, a third body in a hierarchical triple system, as proposed for a small population of eccentric Galactic MSPs with low-mass He white dwarf companions \citep[see][]{grunthal2024triple}. If the tertiary remains bound on a wide orbit, its presence could be revealed by high-order spin-frequency derivatives in future timing analyses \citep[e.g.][]{dutta2025three}.

\subsection{A hint of a hidden history in \textomega~Cen}

Several aspects observed in \textomega~Cen fail to be explained by the currently observed parameters, particularly the low encounter rate per binary and low stellar density: the large fraction of isolated MSPs, the over-abundance of black widow systems, and the large eccentricity of N. Even accepting that the isolated MSP may arise from the black widow population, the conundrum still remains. In this context, it is plausible that a fraction of the MSPs in \textomega~Cen formed in a denser environment than that observed today ($>10^{4}$\,pc$^{-3}$), which would solve the issues discussed above in a natural way.

It is not unreasonable to assume that the internal environment of \textomega~Cen has evolved over time. \textomega~Cen exhibits the greatest stellar population diversity among Galactic GCs and is widely considered to be the remnant nucleus of a former dwarf galaxy. Such diversity may reflect past episodes of star formation, mergers, or dynamical evolution associated with its progenitor system(s) \citep{calamida2020merger}. Furthermore, nuclear clusters in low-mass galaxies (possibly including \textomega~Cen in the past) are thought to form, at least in part, through the in-spiral of globular clusters toward the galactic centre \citep{neumayer2020clusters,fahrion2021diversity}, providing an additional pathway to the structural and dynamical complexity observed in \textomega~Cen.

\section{Conclusions}

In this work, we present the most up-to-date study of the pulsar population of \textomega~Cen. We discovered pulsar S in a recent MeerKAT UHF observation, a new isolated MSP whose localised position has been determined. We searched the APSUSE beams centred on ATCA radio sources in the 21 March 2021 MeerKAT L-band observation to a greater sensitivity than achieved in previous analyses \citep{chen2021mosaic}, but we did not find any pulsed signals.

We derived phase-connected timing solutions for G, H, and K, obtaining the first measurements of their $\dot P_\mathrm{s}$ values and a highly significant detection of $\dot P_\mathrm{b}$ in K. We updated the orbital parameters of I, L, N, and Q via non-phase-connected timing analyses. The revised solutions reveal substantially longer orbital periods for I, N, and Q than previously reported by \citet{chen2023omcen}. Overall, I is an unusual pulsar with a long spin period and a light companion mass, which is only comparable to the Galactic system PSR J1502$-$6752; whereas N has a significant eccentricity, measured here for the first time. Five systems (B, G, H, K, and L) are identified as MSPs in black widows, whereas N and Q are likely to host He white-dwarf companions.

We interpreted the observed $\dot P_\mathrm{b}$ in the timed binaries as arising from binary dynamics instead of the cluster potential. We also measured $\ddot\nu\sim10^{-25}$--$10^{-26}$~Hz\,s$^{-2}$ for pulsars A and H at the 3--5\textsigma~level. While their magnitude would be comparable to what is expected from nearby stars, the measurements are still of low significance and could arise from unmodelled timing noise. 

We used the $\dot P_\mathrm{s}$ measurements of A--E, G, H, and K to constrain the LOS accelerations due to the cluster gravitational potential. All pulsars remain consistent with the LOS acceleration limits predicted by a King-model potential, adopting the updated central velocity dispersion reported by \citet{haberle2025dispersion}. The accelerations also place a 90\% confidence limit of $M_\mathrm{BH}<8\times10^{4}$\,M$_\odot$ on a central IMBH or dark mass. This result is consistent with (although less constraining than) previous limits derived from optical observations and dynamical studies.

We used the newly derived timing solutions for G, H, and K to fold $\gamma$-ray photons from the Fermi-LAT source associated with the cluster, but achieved no individual detection. From the non-detection of pulsed emission from A–E, G, H, and K, we infer that none of these pulsars contribute more than 10\% of the total $\gamma$-ray flux associated with the cluster. Hence, if the $\gamma$-ray flux indeed originates in the pulsar population, the emission might be dominated by pulsars for which we currently do not have a timing solution; alternatively, the emission might originate from the collective emission of several pulsars.

We identified several Chandra X-ray sources as potential counterparts of G, H, K, and L, respectively. Folding the NICER X-ray photon data yielded no pulsed detections. We placed an upper limit of $5.3\times10^{-15}$\,erg\,cm$^{-2}$\,s$^{-1}$ on the X-ray flux for all pulsars from a 2001 XMM-Newton observation \citep{Gendre2003A&A...400..521G}, consistent with previously reported values \citep{henleywillis2018MNRAS.479.2834H}.

The total number of observed MSPs in \textomega~Cen (19) is broadly within the expectations, based on its total stellar encounter rate when compared with other well-studied globular clusters. It is also broadly consistent with its $\gamma$-ray luminosity. However, the population shows very unusual characteristics. The high fractions of isolated MSPs and black widow systems are difficult to reconcile with the low encounter rate per binary observed in the present day. This is in line with the observed excess of X-ray binaries observed in \textomega~Cen in comparison to expectations \citep{deMenezes2023dynamics}. The current encounter rate per binary is also too low to explain the eccentricity of N, a phenomenon also seen in other mildly eccentric MSPs in low-density clusters. 

These findings indicate that present-day encounter rate parameters alone are not reliable predictors of the MSP population in the low-density cluster of \textomega~Cen. We conclude that either there are alternative processes leading to the formation of compact binaries and isolated pulsars or that a significant fraction of the \textomega~Cen MSP population formed in a denser environment ($>10^{4}$\,pc$^{-3}$) than the one observed today. Together with the reported high stellar population diversity in this cluster, these findings point to a complex evolutionary history for \textomega~Cen that remains unresolved.

\section*{Data availability}

Timing and folding solutions for all \textomega~Cen pulsars are available for download at \url{https://doi.org/10.5281/zenodo.23336730}. Scripts used in this publication are available at \url{https://github.com/mcbernadich/pulsar_orbit_solver}.

\begin{acknowledgements}

MCiB has been funded by the INAF Large Grant 2022 “GCjewels” (P.I. Andrea Possenti) approved with the Presidential Decree 30/2022. AP, FA, RN, AC, AC, and MB also benefited of the same "GCjewels" grant. AP was also supported in part by the “Italian Ministry of Foreign Affairs and International Cooperation”, grant number ZA23GR03, under the project "RADIOMAP- Science and technology pathways to MeerKAT+: the Italian and South African synergy".
This work was supported by the Regione Autonoma della Sardegna, under Regional Law n.7 of August 7, 2007 "Promozione della Ricerca Scientifica e dell'Innovazione Tecnologica in Sardegna" (Programma Mobilità Giovani Ricercatori, CUP: F74I19000180002).
FA acknowledges that part of the research activities described in this paper were carried out with the contribution of the NextGenerationEU funds within the National Recovery and Resilience Plan (PNRR), Mission 4 – Education and Research, Component 2 – From Research to Business (M4C2), Investment Line 3.1 – Strengthening and creation of Research Infrastructures, Project IR0000034 – ‘STILES -Strengthening the Italian Leadership in ELT and SKA’. LV acknowledges financial support from the Dean’s Doctoral Scholar Award from the University of Manchester and partial support from NSF grant AST-1816492.
VVK acknowledges continuing support from the Max Planck Society and financial support from the European Research Council (ERC) starting grant ``COMPACT" (Grant agreement number 101078094).
Work at NRL is supported by NASA.

The MeerKAT telescope is operated by the South African Radio Astronomy Observatory (SARAO), which is a facility of the National Research Foundation, an agency of the Department of Science and
Innovation. SARAO acknowledges the ongoing advice and calibration of GPS systems by the National Metrology Institute of South Africa (NMISA) and the time space reference systems department
of the Paris Observatory.
TRAPUM observations used the FBFUSE and APSUSE computing clusters for beamforming, data acquisition, storage and analysis. These instruments were funded, developed and installed by the Max-Planck-Institut für Radioastronomie and the Max-Planck Gesellschaft.
Murriyang, CSIRO’s Parkes radio telescope, is part of the Australia Telescope National Facility (\url{https://ror.org/05qajvd42}), which is funded by the Australian Government for operation as a National Facility managed by CSIRO. We acknowledge the Wiradjuri people as the Traditional Owners of the Observatory site.
The INAF - OAC computer cluster used in this work has been acquired within a project aimed to enhance the Sardinia Radio Telescope (SRT). The Enhancement of the SRT for the study of the Universe at high radio frequencies is financially supported by the National Operative Program (Programma Operativo Nazionale - PON) of the Italian Ministry of University and Research "Research and Innovation 2014-2020", Notice D.D. 424 of 28/02/2018 for the granting of funding aimed at strengthening research infrastructures, in implementation of the Action II.1 - Project Proposal PIR01\_00010.
This work made significant use of the Fornax Science Console, which is part of the NASA Astrophysics cloud-based Fornax Initiative jointly developed by Goddard Space Flight Center’s Astrophysics Projects Division (GSFC/ApPD) and the astrophysics archives—the High Energy Astrophysics Science Archive Research Center (HEASARC), the Infrared Science Archive (IRSA), and the Mikulski Archive for Space Telescopes (MAST).

The Fermi-LAT Collaboration acknowledges generous ongoing support from a number of agencies and institutes that have supported both the development and the operation of the LAT as well as scientific data analysis.
These include the National Aeronautics and Space Administration and the
Department of Energy in the United States, the Commissariat \`a l'Energie Atomique and the Centre National de la Recherche Scientifique / Institut National de Physique Nucl\'eaire et de Physique des Particules in France, the Agenzia Spaziale Italiana and the Istituto Nazionale di Fisica Nucleare in Italy, the Ministry of Education, Culture, Sports, Science and Technology (MEXT), High Energy Accelerator Research Organization (KEK) and Japan Aerospace Exploration Agency (JAXA) in Japan, and the K.~A.~Wallenberg Foundation, the Swedish Research Council and the
Swedish National Space Board in Sweden.

Additional support for science analysis during the operations phase is gratefully acknowledged from the Istituto Nazionale di Astrofisica in Italy and the Centre National d'\'Etudes Spatiales in France. This work performed in part under DOE 
Contract DE-AC02-76SF00515.

This work has made use of Singularity version 3.11. PRESTO and PSRCHIVE software used in this work were installed in Singularity containers provided by AR.
This research makes use of the SciServer science platform (www.sciserver.org) for the analysis of the XMM observations.
SciServer is a collaborative research environment for large-scale data-driven science. It is being developed at, and administered by, the Institute for Data Intensive Engineering and Science at Johns Hopkins University. SciServer is funded by the National Science Foundation through the Data Infrastructure Building Blocks (DIBBs) program and others, as well as by the Alfred P. Sloan Foundation and the Gordon and Betty Moore Foundation.
This research has made use of data, software and/or web tools obtained from the High Energy Astrophysics Science Archive Research Center (HEASARC), a service of the Astrophysics Science Division at NASA/GSFC and of the Smithsonian Astrophysical Observatory's High Energy Astrophysics Division

We give our sincere thanks to the internal MPIfR referee and the anonymous A\&A reviewer for their feedback on this manuscript, which has positively impacted its quality.

\end{acknowledgements}

\bibliography{omcen.bib}

\begin{appendix}

\onecolumn

\section{Full timing parameters tables}

Below, we list the full timing and orbital parameters from the solutions described in Section~\ref{timing_section}. Table~\ref{timing_isolated_appendix} lists the timing parameters for the isolated pulsars A, C, D, and E. Table~\ref{timing_binary_appendix} lists the timing parameters for the binary pulsars B, G, H, and K. Table~\ref{orbital_binary_appendix} lists the timing parameters from the jumped timing solutions (orbital solutions) for binary pulsars I, L, N, and Q.

\begin{table}[h!]
\caption[]{\label{timing_isolated_appendix}Data reduction, spin, astrometric, and derived parameters for isolated pulsars with timing solutions.}
\centering
\begin{tabular}{lcccc}
\hline
\hline \\[-1.5ex]
PSR & J1326$-$4728A & J1326$-$4728C & J1326$-$4728D & J1326$-$4728E \\
\hline \\[-1.5ex]
\multicolumn{3}{l}{ToA integration time} &  \\
\hline \\[-1.5ex]
MeerKAT L band (mins) & 30  & 30  & 30  & 30  \\
MeerKAT S band (mins) & 30  & 30  & 30  & 30  \\
MeerKAT UHF (mins) & 30  & 30  & 30  & 30  \\
Murriyang UWL (mins) & $>180$ & $>180$  & $>180$  & $>180$ \\
\hline \\[-1.5ex]
\multicolumn{2}{l}{Data reduction parameters} &  &  &  \\
\hline \\[-1.5ex]
Ephem. & DE440 & DE440 & DE440 & DE440 \\
Time eph. & IF99 & IF99 & IF99 & IF99 \\
Units & TCB & TCB & TCB & TCB \\
T2CMETHOD  & IAU2000B & IAU2000B & IAU2000B & IAU2000B \\
EPHVER & 5 & 5 & 5 & 5 \\
Clock  corr. & TT(TAI) & TT(TAI) & TT(TAI) & TT(TAI) \\
First ToA & 58962 & 58961 & 58941 & 58941 \\
Last ToA & 60815 & 60885 & 60885 & 60815 \\
\# ToAs & 275 & 250 & 240 & 257 \\
RMS (\textmu s) & 6.899 & 24.476 & 17.247 & 8.763 \\
$\chi^2_\mathrm{r}$ & 1.1363 & 1.1337 & 1.1002 & 0.9595 \\[0.5ex]
\hline \\[-1.5ex]
\multicolumn{2}{l}{Spin and astrometric parameters} &  &  &  \\
\hline \\[-1.5ex]
RA (J2000) & 13$^\mathrm{h}$26$^\mathrm{m}$39.66801(4)$^\mathrm{s}$ & 13$^\mathrm{h}$26$^\mathrm{m}$55.22016(14)$^\mathrm{s}$ & 13$^\mathrm{h}$26$^\mathrm{m}$32.71192(11)$^\mathrm{s}$ & 13$^\mathrm{h}$26$^\mathrm{m}$42.67711(4)$^\mathrm{s}$ \\
DEC (J2000) & $-$47:30:11.6668(5) & $-$47:30:11.774(2) & $-$47:28:40.0746(15) & $-$47:27:24.0232(7) \\
$\mu_{\mathrm{RA}}$ (mas\,yr$^{-1}$) & $-3.8\pm0.3$ & $-4.1\pm0.9$ & $-4.3\pm0.7$ & $-4.0\pm0.3$ \\
$\mu_\textrm{DEC}$ (mas\,yr$^{-1}$) & $-5.8\pm0.4$ & $-9.7\pm1.5$ & $-8.3\pm1.1$ & $-7.1\pm0.5$ \\
$\nu$ (Hz) & 243.380879991247(8) & 145.605770154657(14) & 218.396237332763(15) & 237.658566377299(8) \\
$\dot\nu$ (Hz\,s$^{-1}$) & $-1.61915(14)\times10^{-15}$ & $-2.152(3)\times10^{-16}$ & $1.9644(3)\times10^{-15}$ & $-9.1884(16)\times10^{-16}$ \\
$\ddot\nu$ (Hz\,s$^{-2}$) & $(5.0\pm1.0)\times10^{-26}$ & $<8\times10^{-26}$ & $<8\times10^{-26}$ & $<4\times10^{-26}$ \\
DM$_0$ (pc\,cm$^{-3}$) & 100.3267(3) & 100.6680(12) & 96.5467(8) & 94.3966(4) \\
DM$_1$ (pc\,cm$^{-3}$\,yr$^{-1}$) & $(2.34\pm0.19)\times10^{-3}$ & $(4.3\pm0.7)\times10^{-5}$ & $(1.9\pm4.6)\times10^{-4}$ & $(5.4\pm2.8)\times10^{-4}$ \\
Epoch (MJD) & 59923 & 59923 & 59923 & 59923 \\
PX (mas) & 0.191 & 0.191 & 0.191 & 0.191 \\[0.5ex]
\hline \\[-1.5ex]
\multicolumn{2}{l}{Derived parameters} & & \\
\hline \\[-1.5ex]
$P_\textrm{s}$ (ms) & 4.10878619567800(14) & 6.8678596935948(6) & 4.5788334644078(3) & 4.20771704232379(15) \\
$\dot P_\textrm{s}$ (s\,s$^{-1}$) & $2.7335(2)\times10^{-20}$ & $1.0150(16)\times10^{-20}$ & $-4.1184(7)\times10^{-20}$ & $1.6268(3)\times10^{-20}$ \\
$\dot P_\textrm{Sh}$ (s\,s$^{-1}$) & $(2.6\pm0.3)\times10^{-21}$ & $(1.0\pm0.3)\times10^{-20}$ & $(5.3\pm1.1)\times10^{-21}$ & $(3.7\pm0.4)\times10^{-21}$ \\
$\dot P_\textrm{G}$ (s\,s$^{-1}$) & $(-1.4\pm0.3)\times10^{-21}$ & $(-2.4\pm0.5)\times10^{-21}$ & $(-1.6\pm0.3)\times10^{-21}$ & $(-1.5\pm0.4)\times10^{-21}$ \\
$a_\textrm{LOS,c}$ (m\,s$^{-2}$) & $<1.91(3)\times10^{-9}$ & $<(1.0\pm1.2)\times10^{-10}$ & $<-2.94(8)\times10^{-9}$ & $<1.00(4)\times10^{-9}$ \\
\hline
\hline
\end{tabular}
\tablefoot{For the isolated pulsars we implemented full time integration of MeerKAT UWL observations, deriving a single ToA per observation and sub-band. Hence, ToA integration times are always 180 minutes or longer.\\
Models for which only an upper limit on $\ddot\nu$ is reported include only $\nu$ and $\dot\nu$ in the fit. The quoted upper limits on $\ddot\nu$ are derived from the $3\sigma$ uncertainties obtained by temporarily including $\ddot\nu$ as a free parameter in the timing model.}

\end{table}

\begin{table}
\caption[]{\label{timing_binary_appendix} Data reduction, spin, astrometric, orbital, and derived parameters for binary pulsars with timing solutions.}
\centering
\begin{tabular}{lcccc}
\hline
\hline \\[-1.5ex]
PSR & J1326$-$4728B & J1326$-$4728G & J1326$-$4728H & J1326$-$4728K \\
\hline \\[-1.5ex]
\multicolumn{3}{l}{ToA integration time} &  \\
\hline \\[-1.5ex]
MeerKAT L band (mins) & 2 & 5 & 20  & 2.5 \\
MeerKAT S band (mins) & 2 & 5 & 120 & 10/2.5$^\mathrm{a}$ \\
MeerKAT UHF (mins) & 2 & 5    & 20  & 2.5 \\
Murriyang UWL (mins) & 30.0 & 30.0 & 60.0  & 30.0 \\
\hline \\[-1.5ex]
\multicolumn{2}{l}{Data reduction parameters} &  &  &  \\
\hline \\[-1.5ex]
Ephem. & DE440 & DE440 & DE440 & DE440 \\
Units & TCB & TCB & TCB & TCB \\
Time eph. & IF99 & IF99 & IF99 & IF99 \\
T2CMETHOD  & IAU2000B & IAU2000B & IAU2000B & IAU2000B \\
EPHVER & 5 & 5 & 5 & 5 \\
Clock  corr. & TT(TAI) & TT(TAI) & TT(TAI) & TT(TAI) \\
First ToA & 58941 & 58962 & 58962 & 58962 \\
Last ToA & 60885 & 60885 & 60815 & 60885 \\
\# ToAs & 1526 & 1082 & 435 & 1305 \\
Binary model & ELL1 & ELL1 & ELL1 & ELL1 \\
RMS (\textmu s) & 26.186 & 28.967 & 11.877 & 28.659 \\
$\chi^2_\mathrm{r}$ & 3.7410 & 1.9655 & 1.2351 & 1.5121 \\[0.5ex]
\hline \\[-1.5ex]
\multicolumn{2}{l}{Spin and astrometric parameters} &  &  &  \\
\hline \\[-1.5ex]
RA (J2000) & 13$^\mathrm{h}$26$^\mathrm{m}$49.56780(7)$^\mathrm{s}$ & 13$^\mathrm{h}$26$^\mathrm{m}$37.26302(10)$^\mathrm{s}$ & 13$^\mathrm{h}$26$^\mathrm{m}$44.08334(6)$^\mathrm{s}$ & 13$^\mathrm{h}$26$^\mathrm{m}$38.26700(11)$^\mathrm{s}$ \\
DEC (J2000) & $-$47:29:24.9153(9) & $-$47:29:42.7712(13) & $-$47:28:55.0721(8) & $-$47:27:40.5446(15) \\
$\mu_{\mathrm{RA}}$ (mas\,yr$^{-1}$) & $-2.4\pm0.5$ & $-2.7\pm0.7$ & $-3.7\pm0.4$ & $-2.9\pm0.7$ \\
$\mu_\textrm{DEC}$ (mas\,yr$^{-1}$) & $-9.8\pm0.7$ & $-6.8\pm1.0$ & $-7.0\pm0.6$ & $-5.8\pm1.0$ \\
$\nu$ (Hz) & 208.686833369211(11) & 302.628331143870(19) & 396.75802442509(2) & 212.052881007660(13) \\
$\dot\nu$ (Hz\,s$^{-1}$) & $2.3664(3)\times10^{-15}$ & $-2.5401(4)\times10^{-15}$ & $-6.2769(3)\times10^{15}$ & $4.082(4)\times10^{-16}$ \\
$\ddot\nu$ (Hz\,s$^{-2}$) & $<6\times10^{-26}$ & $<1.2\times10^{-25}$ & $(7.2\pm2.4)\times10^{-26}$ & $<9\times10^{-26}$ \\
DM$_0$ (pc\,cm$^{-3}$) & 100.2859(6) & 99.7431(8) & 98.1715(5) & 94.7892(7)\\
DM$_1$ (pc\,cm$^{-3}$\,yr$^{-1}$) & $(2.3\pm4.6)\times10^{-4}$ & $(-1.9\pm0.5)\times10^{-3}$ & $(-1.4\pm0.3)\times10^{-3}$ & $(-4.2\pm4.2)\times10^{-4}$ \\
Epoch (MJD) & 59923 & 59923 & 59923 & 59923 \\
PX (mas) & 0.191 & 0.191 & 0.191 & 0.191 \\[0.5ex]
\hline \\[-1.5ex]
\multicolumn{2}{l}{Orbital parameters} &  &  \\
\hline \\[-1.5ex]
$P_\textrm{b}$ (days) & 0.08961120476(14) & 0.10875952303(14) & 0.13569376184(19) & 0.09387146340(5) \\
$\dot P_\textrm{b}$ (s\,s$^{-1}$) & $(3.8\pm0.5)\times10^{-12}$ & $(2.1\pm0.5)\times10^{-12}$ & $(-1.7\pm0.7)\times10^{-12}$ & $6.24(16)\times10^{-12}$ \\
$x$ (ls) & 0.0214545(10) & 0.0322022(14) &  0.0218766(9) & 0.0679499(19) \\
$T_\mathrm{A}$ (MJD) & 59923.1256581(9) & 59923.4075032(9) & 59923.7586019(11) & 59923.7258053(3) \\
$\eta$ & $(2.9\pm9.3)\times10^{-5}$ & $(2.3\pm0.8)\times10^{-4}$ & $(-3.1\pm8.2)\times10^{-5}$ & $(4.6\pm4.4)\times10^{-5}$ \\
$\kappa$ & $(-5.5\pm0.9)\times10^{-4}$ & $(-2.1\pm0.8)\times10^{-4}$ & $(-1.2\pm0.8)\times10^{-4}$ & $(-6.6\pm3.6)\times10^{-5}$ \\
\hline \\[-1.5ex]
\multicolumn{2}{l}{Derived parameters} & & \\
\hline \\[-1.5ex]
$P_\textrm{s}$ (ms) & 4.7918691555916(2) & 3.3043832883069(2) & 2.52042791434154(13) & 4.7158048277773(3) \\
$\dot P_\textrm{s}$ (s\,s$^{-1}$) & $-5.4337(6)\times10^{-20}$ & $2.7736(5)\times10^{-20}$ & $3.9874(2)\times10^{-20}$ & $-9.077(8)\times10^{-21}$ \\
$\dot P_\textrm{Sh}$ (s\,s$^{-1}$) & $(6.5\pm0.9)\times10^{-21}$ & $(2.4\pm0.6)\times10^{-21}$ & $2.1(3)\times10^{-21}$ & $(2.7\pm0.8)\times10^{-21}$ \\
$\dot P_\textrm{G}$ (s\,s$^{-1}$) & $(-1.7\pm0.3)\times10^{-21}$ & $(-1.1\pm0.2)\times10^{-21}$ & $(-8.7\pm1.7)\times10^{-21}$ & $(-1.6\pm0.3)\times10^{-21}$  \\
$a_\textrm{LOS,c}$ (m\,s$^{-2}$) & $<-3.70(6)\times10^{-9}$ & $<2.41(6)\times10^{-9}$ & $<4.60(4)\times10^{-9}$ & $<(-6.4\pm0.5)\times10^{-10}$ \\
$M_\mathrm{c,min}$ (M$_\odot$) & $0.013$ & $0.018$ & $0.010$ & $0.042$ \\
$M_\mathrm{c,med}$ (M$_\odot$) & $0.016$ & $0.021$ & $0.012$ & $0.049$ \\
\hline
\hline
\end{tabular}

\tablefoot{Models for which only an upper limit on $\ddot\nu$ is reported include only $\nu$ and $\dot\nu$ in the fit. The quoted upper limits on $\ddot\nu$ are derived from the $3\sigma$ uncertainties obtained by temporarily including $\ddot\nu$ as a free parameter in the timing model.\\
$M_\textrm{c,min}$ and $M_\textrm{c,med}$ derived from the mass function assuming $M_\mathrm{p}=1.35$\,$M_\odot$ and $i=90,\,60^\circ$).\\
$^\mathrm{a}$Pulsar K presents high variability in observed flux, affecting its detectability at the MeerKAT S-band. For the 21 \& 22 January 2024 observations, it was necessary to integrate over 10 minutes to measure reliable ToAs, while in the 12 February 2024 observation we could maintain a 2.5-minute resolution.}

\end{table}

\begin{table}[h!]
\caption[]{\label{orbital_binary_appendix} Data reduction, spin, astrometric, orbital, and derived parameters for binary pulsars with not-connected orbital solutions.}
\centering
\begin{tabular}{lcccc}
\hline
\hline \\[-1.5ex]
PSR & J1326$-$4728I & J1326$-$4728L & J1326$-$4728N & J1326$-$4728Q \\
\hline \\[-1.5ex]
\multicolumn{2}{l}{Data reduction parameters} &  &  &  \\
\hline \\[-1.5ex]
Ephem. & DE430 & DE430 & DE430 & DE430 \\
Units & TCB & TCB & TCB & TCB \\
Data set & MeerKAT & MeerKAT & MeerKAT & MeerKAT \\
First ToA & 59294 & 59294 & 59294 & 59294 \\
Last ToA & 60884 & 60884 & 60884 & 60717 \\
\# ToAs & 47 & 78 & 43 & 69 \\
Binary model & ELL1 & ELL1 & DD & ELL1 \\
RMS (\textmu s) & 83.241 & 33.305 & 23.345 & 31.351 \\
$\chi^2_\mathrm{r}$ & 0.8360 & 1.3777 & 1.3184 & 0.8344 \\[0.5ex]
\hline \\[-1.5ex]
\multicolumn{2}{l}{Spin and astrometric parameters} &  &  &  \\
\hline \\[-1.5ex]
RA (J2000) & 13$^\mathrm{h}$26$^\mathrm{m}$29.0$^\mathrm{s}$ & 13$^\mathrm{h}$27$^\mathrm{m}$02.8$^\mathrm{s}$ & 13$^\mathrm{h}$26$^\mathrm{m}$49.8$^\mathrm{s}$ & 13$^\mathrm{h}$26$^\mathrm{m}$35$^\mathrm{s}$ \\
DEC (J2000) & $-$47:30:24 & $-$47:26:49 & $-$47:31:25 & $-$47:27:54 \\
$\nu$ (Hz) & 52.7673960(13) & 282.7492572(3) & 145.2721044(14) & 242.123211(7) \\
DM$_0$ (pc\,cm$^{-3}$) & 102.555 & 101.4759 & 102.1 & 95.923 \\
\hline \\[-1.5ex]
\multicolumn{2}{l}{Orbital parameters} &  &  \\
\hline \\[-1.5ex]
$P_\textrm{b}$ (days) & 3.508256(2) & 0.1589288685(7) & 6.3566264(6) & 1.5400249(3) \\
$x$ (ls) & 0.7852(17) & 0.061800(6) & 5.7231(14) & 1.9938(2) \\
$T_\mathrm{A}$ (MJD) & 59293.4485(9) & 59294.704979(4) & ... & 59282.38142(12) \\
$\eta$ & $(-4.6\pm13.2)\times10^{-4}$ & $(1.0\pm1.9)\times10^{-4}$ & ... & $(4.4\pm9.7)\times10^{-5}$ \\
$\kappa$ & $(1.6\pm0.9)\times10^{-3}$ & $(1.4\pm2.0)\times10^{-4}$ & ... & $(5.5\pm18.5)\times10^{-5}$ \\
$T_0$ (MJD) & ... & ... & 59292.3198(12) & ... \\
$\omega$ ($^\circ$) & ... & ... & 241.78(7) & ... \\
$e$ & ... & ... & 0.09282(11) & ...  \\
\hline \\[-1.5ex]
\multicolumn{2}{l}{Derived parameters} & & \\
\hline \\[-1.5ex]
$P_\textrm{s}$ (ms) & 18.9510962(5) & 3.536702482(4) & 6.88363402(7) & 4.13012861(12) \\
$M_\mathrm{c,min}$ (M$_\odot$) & $0.043$ & $0.027$ & $0.232$ & $0.205$ \\
$M_\mathrm{c,med}$ (M$_\odot$) & $0.050$ & $0.031$ & $0.272$ & $0.241$ \\
\hline
\hline
\end{tabular}

\tablefoot{Out of the spin and astrometric parameters, only $\nu$ is a free parameter in the fits. $M_\textrm{c,min}$ and $M_\textrm{c,med}$ derived from the mass function assuming $M_\mathrm{p}=1.35$\,$M_\odot$ and $i=90,\,60^\circ$).}

\end{table}

\clearpage

\section{Details about MeerKAT observations}\label{observations_appendix}

\begin{table*}[h!]
\caption[]{\label{observations_table} Summary of MeerKAT APSUSE observations used in this work.}
\centering
\resizebox{\textwidth}{!}{\begin{tabular}{lccccccccc}
\hline
\hline \\[-1.5ex]
Date & Project & Receiver & Duration & Channels & Sampling    & Array & Tiling & Comments \\
     &         &          & (hours)  & & (\textmu s) &       &        &          \\
\hline \\[-1.5ex]
2021-03-21 & TRAPUM   & L band & 3.8 & 2048/128 & 153.121 & Core  & M, A, P & Searched in \citep{chen2023omcen} \\
2021-03-26 & TRAPUM   & L band & 3.7 & 2048/128 & 153.121 & Core  & M, A, P & Searched in \citep{chen2023omcen} \\
2023-06-27 & MeerTIME & L band & 4.0 & 4096/256 & 76.5607 &  Full & A, Pm & Follow-up \\
2023-08-20 & MeerTIME & L band & 4.0 & 4096/256 & 76.5607 &  Full & A, Pm & Follow-up \\
2023-10-05 & MeerTIME & L band & 4.0 & 4096/256 & 76.5607 &  Full & A, Pm & All beams corrupted, data not used \\
2024-01-21 & MeerTIME & S band & 2.0 & 1024/64 & 18.7246 &  Full & A, Pm & \small{Part of the S-band GC survey \citep{nag2026sband}} \\ 
2024-01-22 & MeerTIME & S band & 2.0 & 1024/64 & 18.7246 &  Full & A, Pm & \small{Starts at the end of previous, 4 hours total}\\
2024-02-12 & MeerTIME & S band & 2.0 & 1024/64 & 74.8983 &  Full & A, Pm & \small{Some beams lost, $>2.7$\,GHz sub-band lost} \\
2025-02-11 & TRAPUM   & UHF    & 3.0 & 4096/256 & 120.471 &  Core & M(0.7), A, Pm & Searched in this work \\
2025-07-28 & TRAPUM   & UHF    & 3.0 & 4096/256 & 240.941 &  Full & M(0.7), A, Pm & $>0.95$\,GHz sub-band lost \\

\hline
\hline
\end{tabular}
}
\tablefoot{PTUSE observation from 14 February, 2024 is not included, as it was only used for pulsar L. Core array: inner 44 antennas. Full array: 55+ antennas. Receiver frequency coverages: L band (856–1712 MHz), S band (S1 configuration, 1968–2843 MHz, \citealt{ebarr2018sband}), and UHF (544–1088 MHz).  The channels indicate channels recorded over the channels after sub-banding. M: MOSAIC coherent beam tiling (0.7: defined beam overlap, if set). A: coherent beams on ATCA sources \citep{dai2023omcen}. P: coherent beams on locations of pulsars A and B \citep{dai2023omcen}. Pm: coherent beams on locations of all Parkes and MeerKAT pulsar discoveries \citep{dai2023omcen,chen2023omcen}.}

\end{table*}

MeerKAT observations were processed on the Filterbanking Beamformer User Supplied Equipment (FBFUSE) cluster for beamforming, and were subsequently stored as search-mode filterbanks (without coherent de-dispersion) in the Accelerated Pulsar Search User Supplied Equipment (APSUSE) \citep[see][]{ebarr2018sband,padmanabh2023mmgps}.

APSUSE does not support coherent de-dispersion, and therefore none of the recorded MeerKAT data products used in this work includes it. TRAPUM is primarily a search-focused project, and hence its observations implemented coherent beam tiling. In contrast, MeerTIME observations were follow-up observations, with the main data products consisting of beams from the Pulsar Timing User Supplied Equipment (PTUSE) backend \citep{bailes2020facility}, which are primarily intended for other projects (e.g. the S-band GC survey; \citealt{nag2026sband}; or the eclipses of pulsars B, G, K, and L). Nonetheless, individual APSUSE beams centred on the known positions of pulsars and ATCA sources were also recorded.

The 12 February 2024 and 28 July 2025 observations suffered partial recording failures, resulting in the loss of a fraction of the observing band in both cases; additionally, during the 12 February 2024 observation, a subset of beams was corrupted and lost entirely. Nevertheless, the remaining usable bandwidth and beams were retained and used in the scientific analysis presented in this work.

\clearpage

\section{Sky coverage and sensitivity of TRAPUM searches}\label{sky_appendix}

Following \citet{chen2023omcen}, the sensitivity of our search survey is estimated using the radiometer equation \citep{dewey1985search},
\begin{equation}
S_\textrm{min}=\frac{\mathrm{S/N}\,\beta\,T_\textrm{sys}}{\varepsilon_\mathrm{FFT}\,G\,\sqrt{n_\mathrm{pol}\,\Delta f\,\Delta t}}\sqrt{\frac{\zeta}{1-\zeta}}\mathrm{.}
\end{equation}
Taking UHF values, a minimum $\mathrm{S/N}=8$; $\beta=1.01$ is the digitization correction factor; $T_\mathrm{sys}=39.5$\,K is the system temperature, including contributions from the sky at UHF (9.0\,K), spillover and atmosphere (6.5\,K), and the receiver (24\,K); $\varepsilon_\mathrm{FFT}=0.7$ is the FFT efficiency \citep{morello2020ffarevisit}; $G=1.92$\,K\,Jy$^{-1}$ is the combined gain of the MeerKAT core array (44 antennas); $n_\mathrm{pol}=2$ is the number of summed polarisations; $\Delta f=500$\,MHz is the effective bandwidth after radio-frequency interference (RFI) excision; $\Delta t=3$\,h is the integration time; and $\zeta=0.1$ is the assumed pulse duty cycle at UHF.

The sky temperature is estimated via \begin{equation}T_\mathrm{sky}=2.725\mathrm{\,K} + 3.6\mathrm{\,K}\times\left(\frac{f}{GHz}\right)^{-2.75}\mathrm{,}\end{equation} where 3.6 K is the sky temperature at the position of \textomega~Cen at $f=1$~GHz, taken from \citep{remazeilles2015survey}. We adopt $f=0.816$\,GHz for the UHF receivers\footnote{\url{https://skaafrica.atlassian.net/wiki/spaces/ESDKB/pages/277315585/MeerKAT+specifications}}.

\begin{figure}[h!]
\centering
 \includegraphics[width=0.85\hsize]{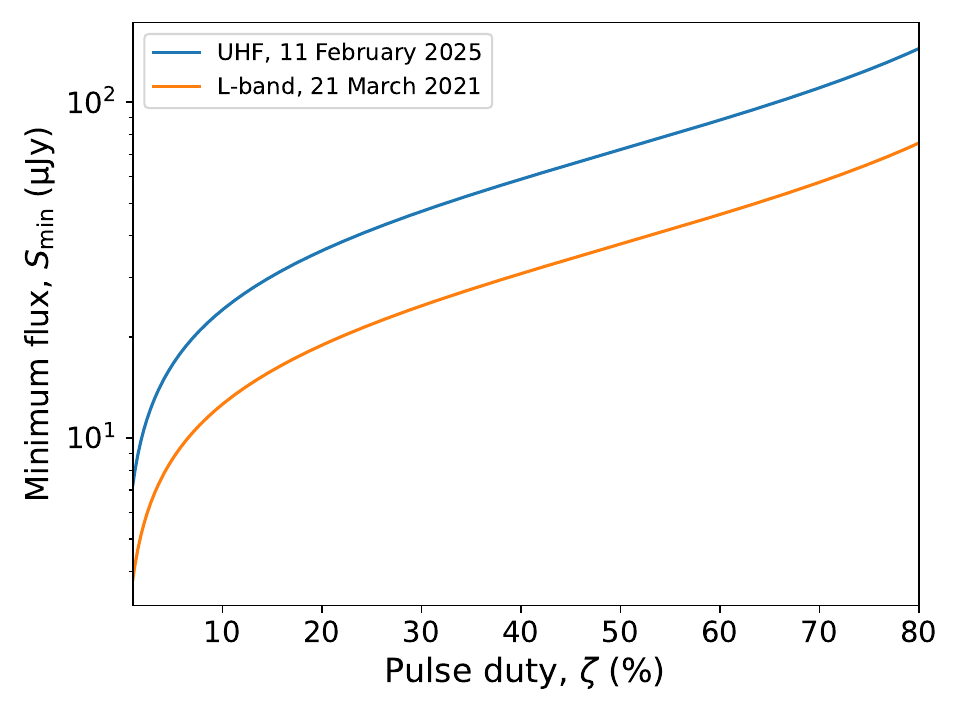}
   \caption{Sensitivity of our (UHF) and the \citet[][L-band]{chen2023omcen} pulsar searches at the beam centres as a function of pulse duty cycle.}
   \label{sensitivity_plot}
\end{figure}

Figure~\ref{sensitivity_plot} shows the evolution of $S_\mathrm{min}$ as a function of $\zeta$ for the UHF configuration stated above, and the L-band configuration ($T_\mathrm{sys}=27$\,K, $\Delta f=640$\,MHz, $\Delta t=4$\,hours, \citealt{chen2023omcen}), assuming no smearing due to DM and scattering effects and optimal source position related to beam sensitivity. The plot shows that at $\zeta=0.1$, the sensitivities are $S_\mathrm{min}\approx20$ for the UHF search and $S_\mathrm{min}\approx10$ for the L-band search. Given a coherent beam tiling overlap of 70\%, positional offsets from beam centre can bring the effective gain down to 70\% in the worst-case scenario, hence increasing $S_\mathrm{min}$ by up to $\sim$30\% at the beam edges.

\clearpage

\begin{figure}[h!]
\centering
 \includegraphics[bb=85 35 640 540,width=0.85\hsize,clip]{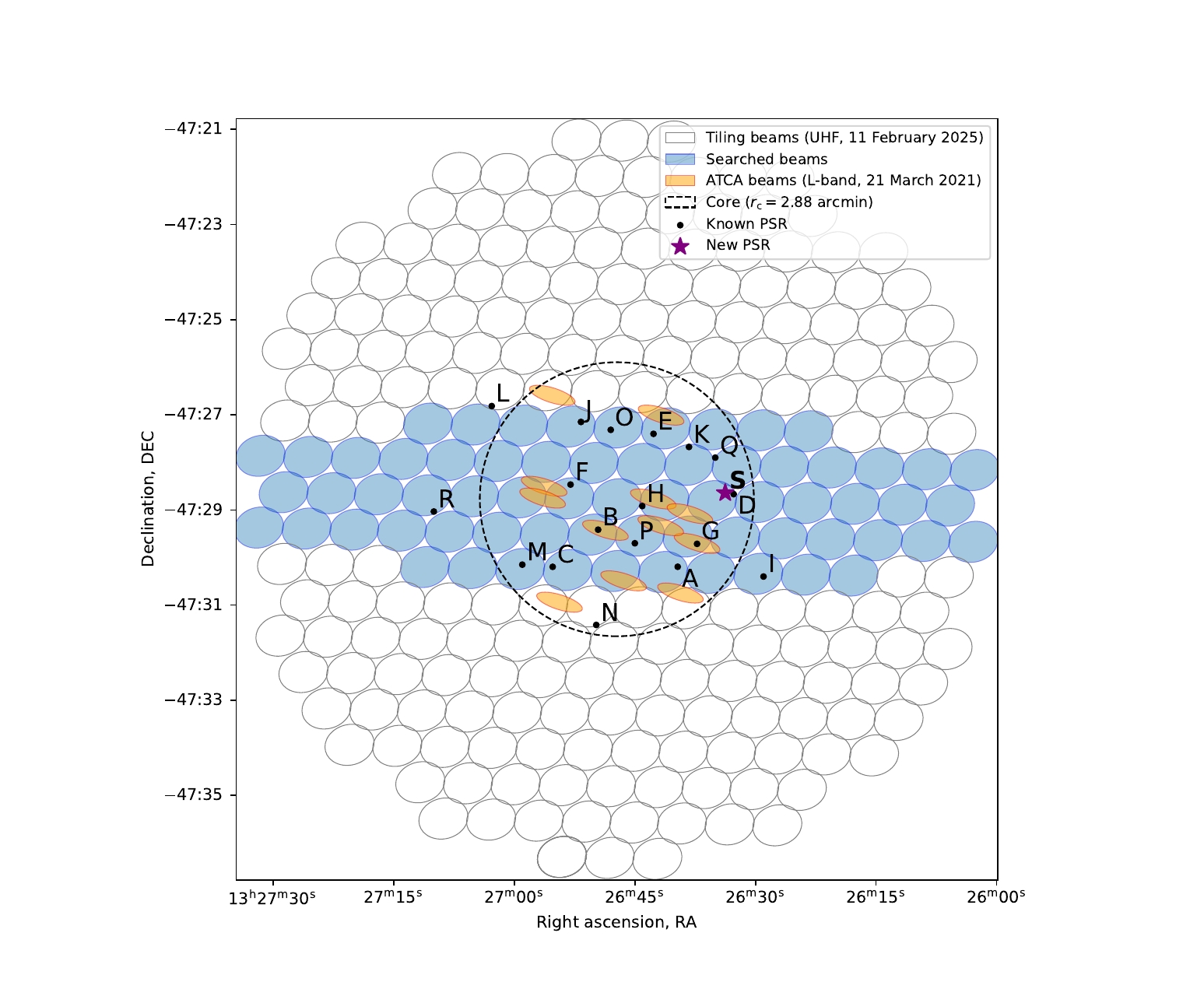}
   \caption{Beams included in the pulsar searches discussed in this work, centred on the optical centre of \textomega~Cen. The beams correspond to the MOSAIC coherent beam tiling generated for the 11 February 2025 observation and to the ATCA beams from the 21 March 2021 observation (searched and inspected beams highlighted with solid colours). The \textomega~Cen core radius, as reported by \citet{baumgardt2018catalogueGCs}, is also shown. Previously known pulsars are indicated, with positions derived from timing or as reported by \citet{chen2023omcen}. Our new UHF discovery, S, is also highlighted at its localised position with a star symbol.}
   \label{beams_map}
\end{figure}

All ATCA beams from the 21 March 2021 observation have been searched. The search for pulsars in the 11 February 2025 observation is an ongoing effort: 66 out of 283 have been searched and their resulting pulsar candidates inspected. This covers most of the core region of \textomega~Cen. Figure~\ref{beams_map} shows the MOSAIC coherent beam tiling of the 11 February 2025 observation and the searched beams.

\clearpage

\section{Derivation of orbital and timing solutions}\label{timing_appendix}

\subsection{Orbital solutions}\label{orbital_solutions_section}

\begin{figure}[h!]
\centering
 \includegraphics[width=0.49\hsize]{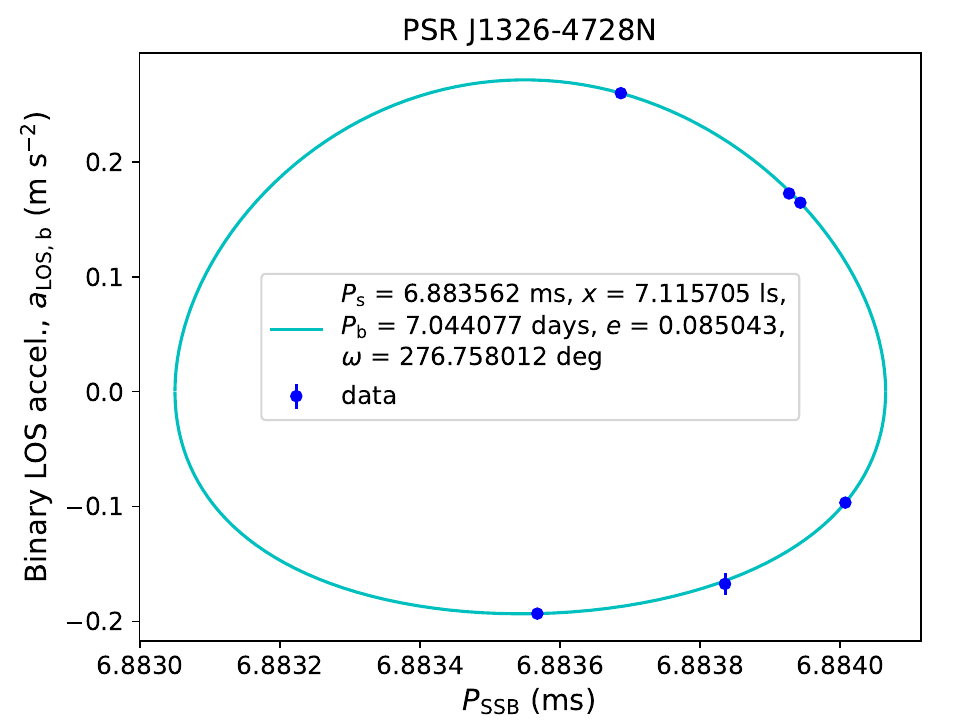}
 \includegraphics[width=0.49\hsize]{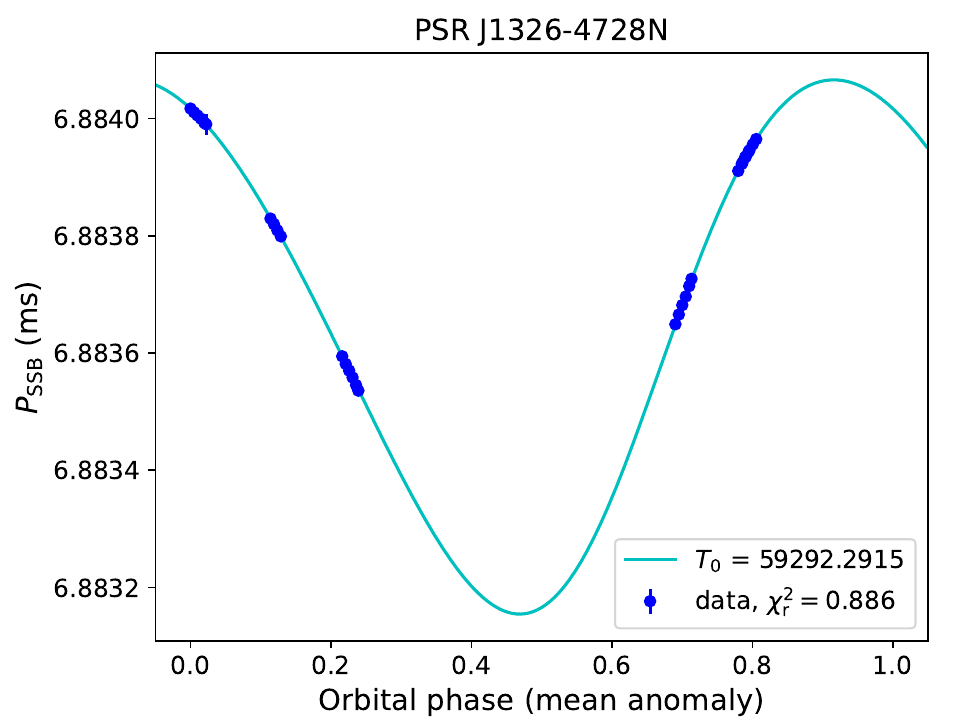}
 \includegraphics[width=\hsize]{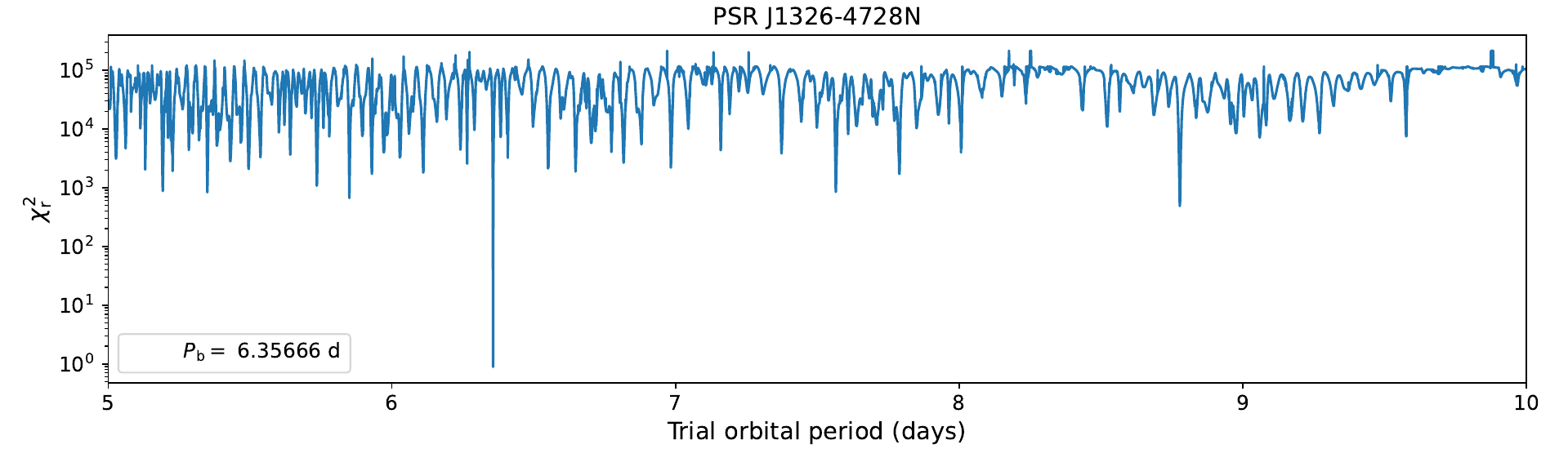}
 \caption{Keplerian orbital fits to the $P_\textrm{SSB}$ and $a_\mathrm{LOS,b}$ measurements of N. \textbf{Top left:} Successful fit of the curve traced in the $P_\textrm{SSB}$–$a_\mathrm{LOS,b}$ plane by a general Keplerian orbit, providing approximate orbital parameters. \textbf{Top right:} Successful fit to the $P_\textrm{SSB}$–orbital phase, folded at the optimal orbital period. \textbf{Bottom:} Quality of the Keplerian fits, expressed as the reduced $\chi^2$ computed across a range of trial orbital periods. Each point corresponds to a general Keplerian fit performed at a given trial period. All plots were generated with our custom code for fitting Keplerian orbits (\url{https://github.com/mcbernadich/pulsar_orbit_solver})}
 \label{N_orbital}
\end{figure}

To obtain preliminary folding solutions for the timing analysis, we first refined the Keplerian orbital solutions for the \textomega~Cen binary pulsars discovered by \citet{chen2023omcen}: G, H, I, K, L, N, and Q. This refinement was necessary to obtain consistent detections across the Murriyang UWL and MeerKAT observations through direct folding of the search-mode data. This section describes this preliminary processing, which preceded the analyses presented in Section~\ref{timing_section}.

The updated orbital solutions were derived from the MeerKAT data set (Table~\ref{observations_table}), owing to its higher sensitivity. We first performed direct refolding using the solutions from \citet{chen2023omcen} with \texttt{dspsr}\footnote{\url{https://github.com/demorest/dspsr}} \citep{vanStraten2011}. This was successful for G, H, K, and L. I, N, and Q were not detected in the newer observations and therefore required additional pulsar searches in their corresponding beams. We searched these beams with \texttt{PRESTO}'s \texttt{accelsearch}, obtaining significant detections in all observations except the S-band observations and, for Q, the 28 July 2025 UHF observation.

Once reliable detections were obtained, we refined the orbital solutions. The pulsar archives were first cleaned of radio-frequency interference using \texttt{clfd}. The folded archives were then divided into segments several minutes long, with the duration chosen according to the brightness of each pulsar. For each segment, we used \texttt{PSRCHIVE}'s \texttt{pdmp}\footnote{\url{https://psrchive.sourceforge.net/manuals/pdmp/}} to measure the barycentric spin period, $P_\textrm{SSB}$, and apparent LOS acceleration due to binary motion, $a_\mathrm{LOS,b}$. We then fitted the resulting $P_\textrm{SSB}$ measurements with Keplerian models using our custom code developed for orbital periodicity searches and Keplerian orbital modelling\footnote{\url{https://github.com/mcbernadich/pulsar_orbit_solver}}.

For G, H, K, and L, this procedure was straightforward owing to the reliable orbital solutions of \citet{chen2023omcen}. In contrast, I, N, and Q required additional steps. We first obtained preliminary Keplerian parameters (orbital period $P_\mathrm{b}$, projected semi-major axis $x$, and eccentricity parameters $e$ and $\omega$) by fitting the curve drawn in the $P_\mathrm{SSB}$--$a_{\mathrm{LOS,b}}$ plane (see the top-left panel of Fig.~\ref{N_orbital}; see \citealt{freire2001determination} for a detailed description).

Using these parameters as priors, we searched for the orbital period by fitting Keplerian models, including the time of periastron $T_0$, to the $P_\mathrm{SSB}$ time series over a grid of trial $P_\mathrm{b}$ values. We selected the period giving the minimum reduced $\chi^2$ (bottom panel of Fig.~\ref{N_orbital}), which ensured orbital coherence between observations separated by many months. Finally, with the best $P_\mathrm{b}$ value, we performed a full Keplerian fit to the $P_\mathrm{SSB}$ time series to obtain the final folding solution (top-right panel of Fig.~\ref{N_orbital}).

The resulting orbital solutions, together with the spin periods of the isolated pulsars discovered by \citet{chen2023omcen}, enabled direct-folding searches in the Murriyang UWL observations. We obtained frequent moderate-S/N detections of G, H, I, and K; occasional low-S/N detections of J and O; and no significant detections of F, L, M, N, P, Q, R, and S. These pulsars were already among the faintest MeerKAT discoveries in \textomega~Cen, making their non-detection with Murriyang UWL unsurprising.

\subsection{Timing solutions}

The folding solutions described above enabled consistent detections in the Murriyang UWL and MeerKAT observations. However, they were not yet suitable starting points for \texttt{TEMPO2} timing fits because they lacked phase connection. Here, we describe the derivation of the preliminary phase-connected timing solutions used as starting points for the \texttt{TEMPO2} fits presented in Section~\ref{timing_section}.

Phase-connected timing solutions were searched jointly using MeerKAT and Murriyang UWL observations for pulsars detectable with both telescopes. Frequency-integrated ToAs were derived every few minutes in each detection using \texttt{PSRCHIVE}'s \texttt{pat}, with high-S/N detections from the 2023 MeerKAT observations as timing templates. For MeerKAT, multiple frequency-integrated ToAs were obtained per observation, including from the S band when detectable. For Murriyang UWL, ToAs were then derived from the lower half of the band ($f<2368$\,MHz) to obtain significant measurements.

We retained only UWL observations with clear detections and low-uncertainty ToAs to ensure reliable time-resolved measurements. These ToAs were combined with the MeerKAT ToAs, with a global jump between the two telescopes, and searched for phase-connected solutions using \texttt{dracula}\footnote{\url{https://github.com/pfreire163/Dracula}}, which explores possible phase-turn combinations between observations to identify a unique solution \citep{freire2018dracula}. The fits were performed with \texttt{Tempo}\footnote{\url{https://tempo.sourceforge.net/}} \citep{nice2015tempo}, the fitting software on which \texttt{dracula} runs, and included the spin frequency $\nu$ and its derivative $\dot\nu$, Keplerian orbital parameters for binaries, and the position parameters RA and DEC.

Phase-connected timing solutions were obtained for G, H, and K, spanning both telescopes from 2020 to 2025. We also attempted phase connection for I, J, and O, which provided frequent, high-S/N detections, but could not obtain a unique solution across the full data span. Pulsar I was particularly promising, with a brightness comparable to G and H, but we were unable to find any timing solution describing the entire data set.

\clearpage

\section{Chandra sources and NICER FoV}\label{chandra_appendix}

\begin{figure}[h!]
\centering
 \includegraphics[width=0.88\hsize]{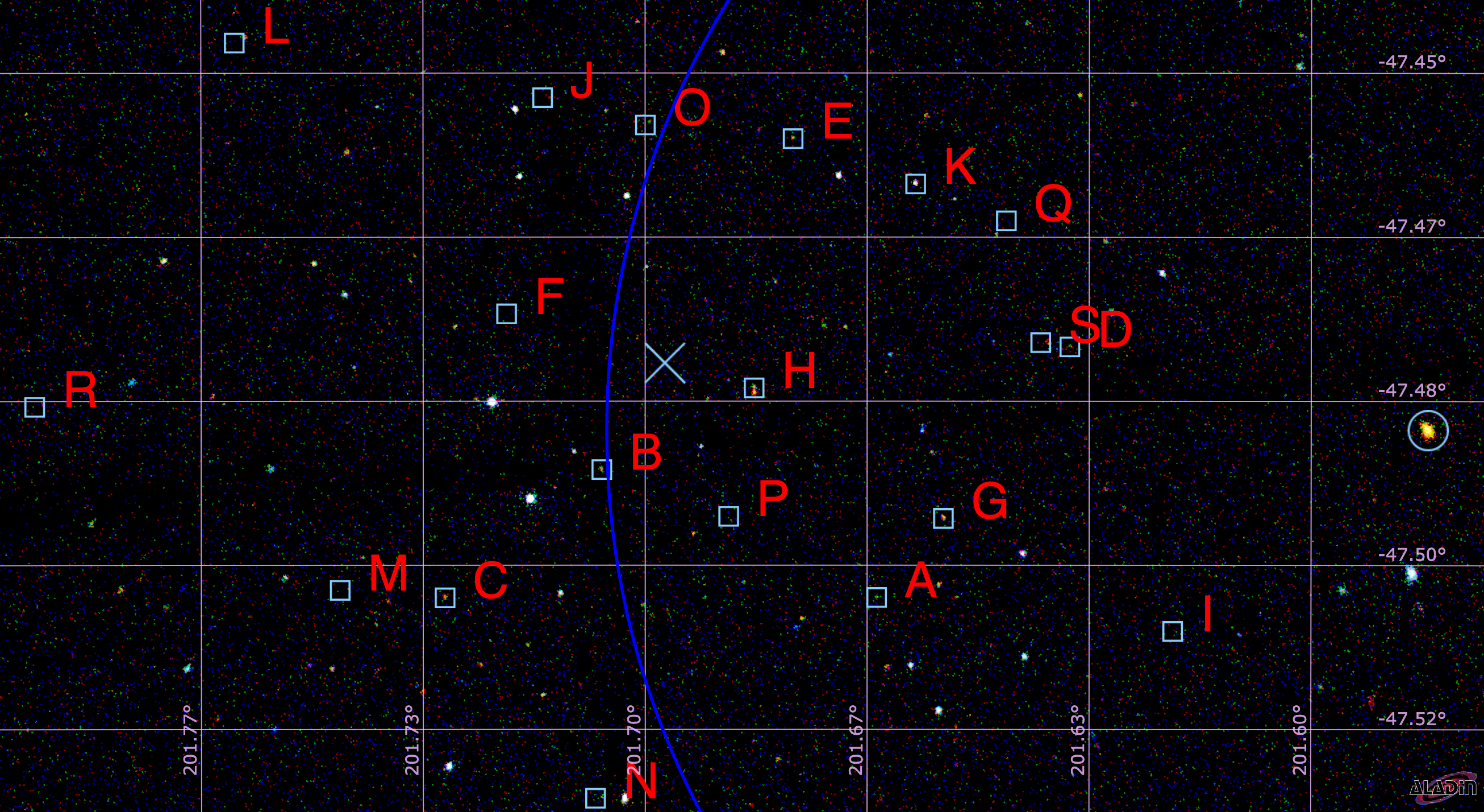}
   \caption{\textomega~Cen pulsar population overlaid on a Chandra X-ray image (via the Aladin Sky Atlas; \citealt{baumann2022aladin}), showing several source coincidences (A, B, C, E, G, H, K and potentially L). The cross marks the cluster centre; the large blue circle shows the NICER FoV analysed here, centred on the quiescent LMXB CXOU 132619.7–472910.8 (cyan; \citealt{rutledge2002transient}).}
   \label{chandra_map}
\end{figure}

\begin{figure}[h!]
\centering
 \includegraphics[width=0.39\hsize]{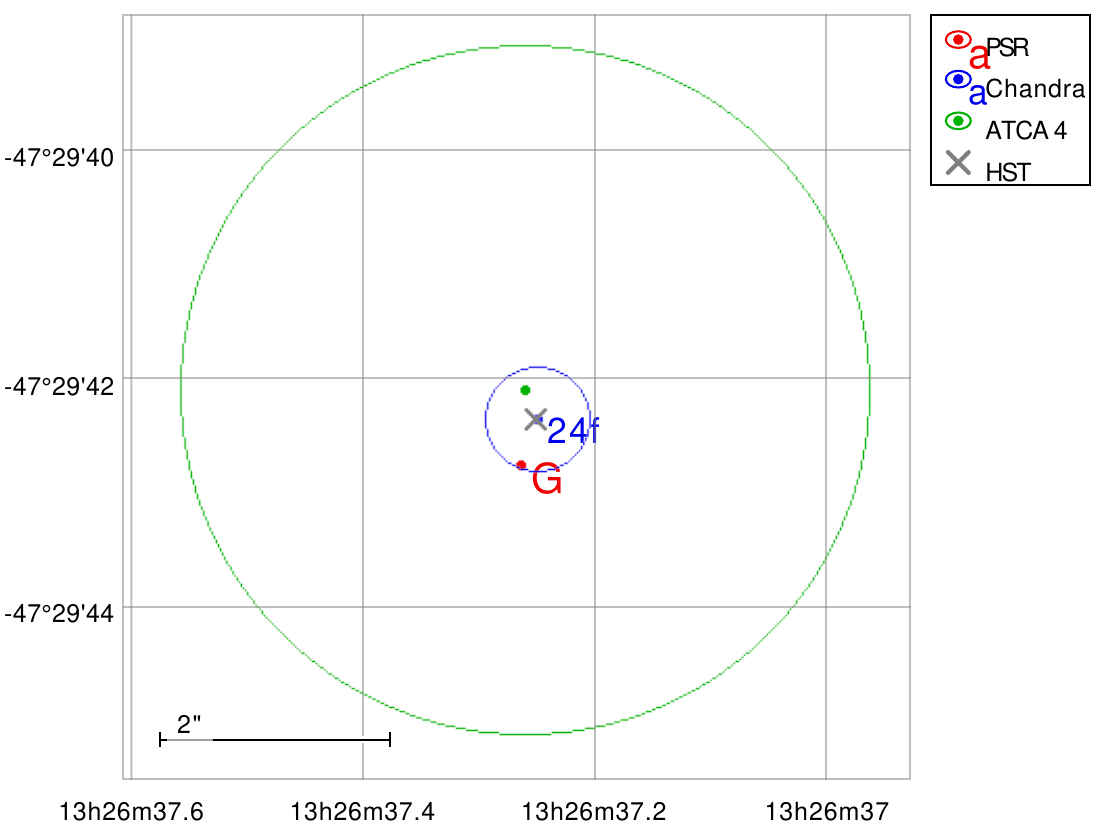}
 \includegraphics[width=0.39\hsize]{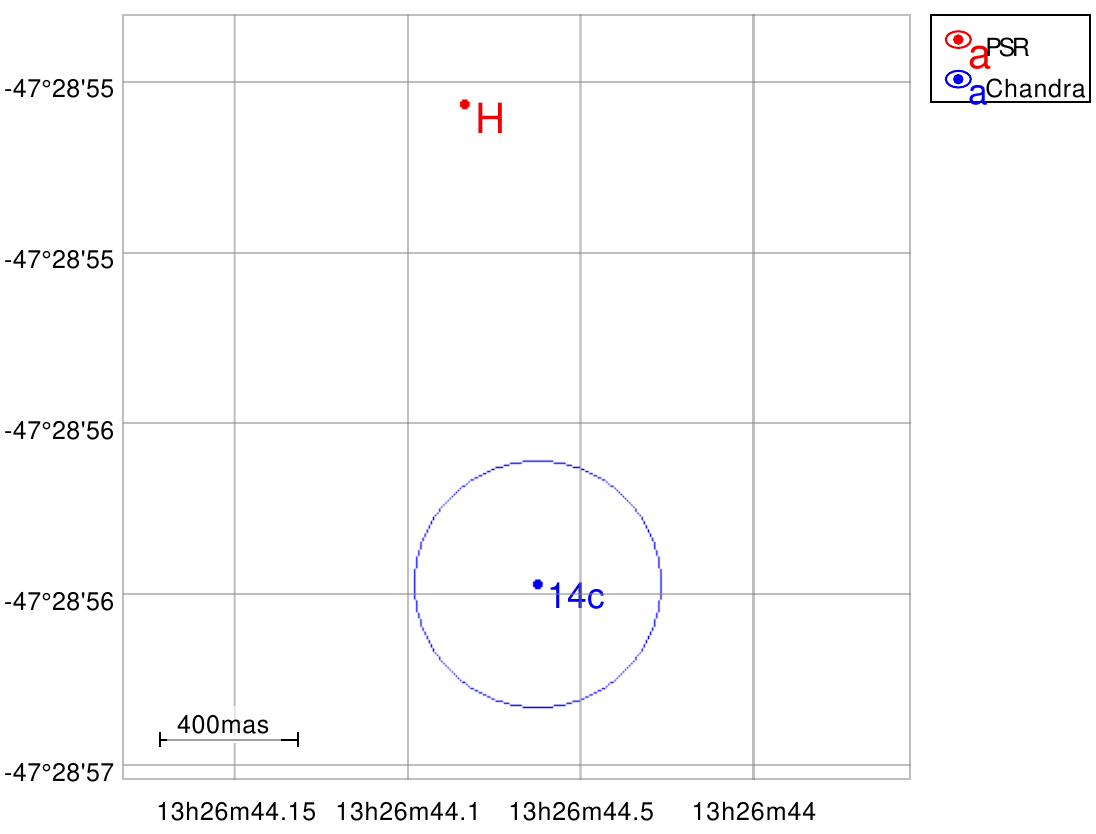}
 \includegraphics[width=0.39\hsize]{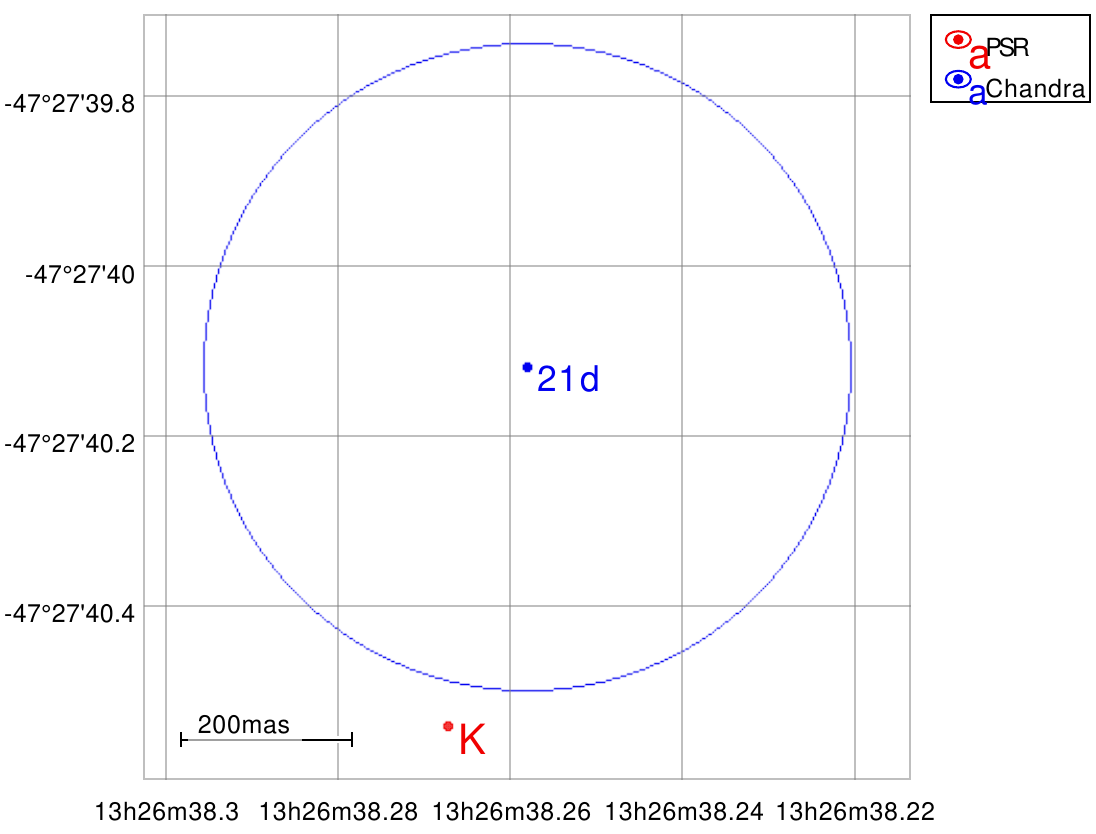}
 \includegraphics[width=0.39\hsize]{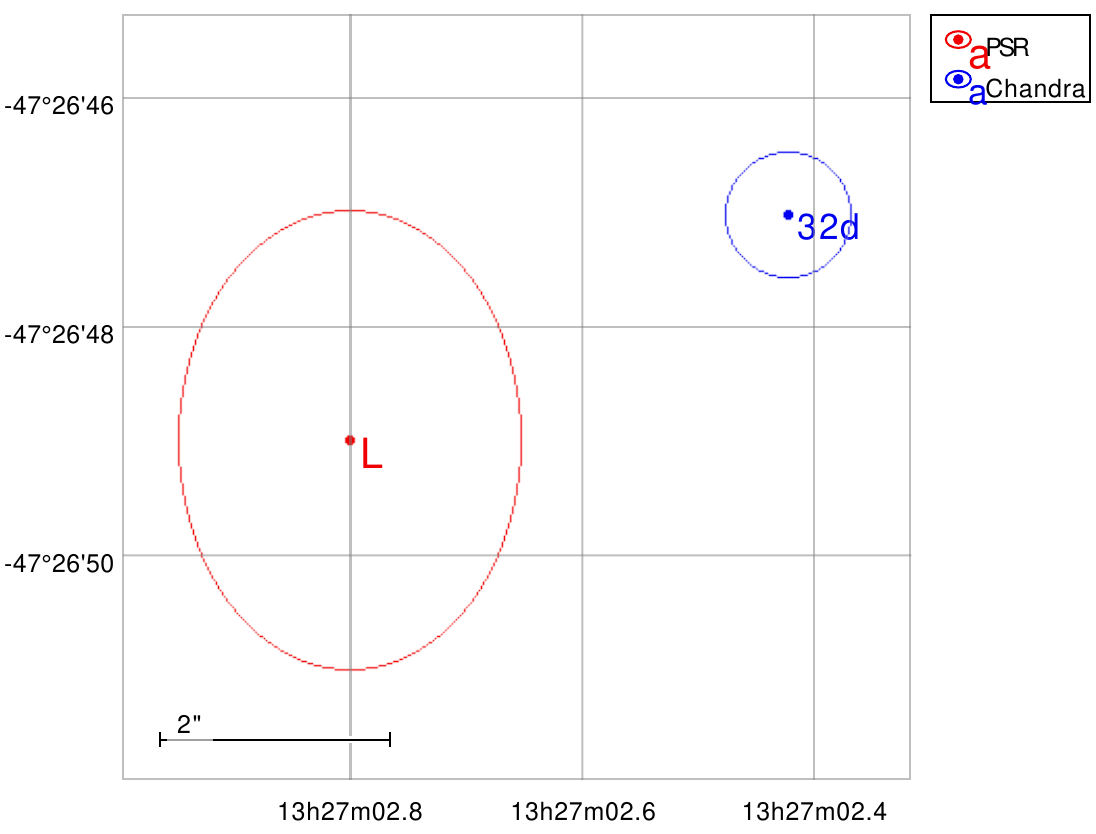}
 \caption{Positions of G, H, K, and L compared with their nearest Chandra X-ray (\citealt{henleywillis2018MNRAS.479.2834H}), ATCA (\citealt{dai2023omcen}), and HST (\citealt{cool2013hst/acs}) counterparts. Timing positions are used for G, H, and K; L uses the localisation from \citealt{chen2023omcen}. Drawn with TOPCAT \citep[\url{http://www.starlink.ac.uk/topcat/},][]{taylor2005topcat}.}
 \label{counterparts_plots}
\end{figure}

\end{appendix}

\end{document}